\documentclass[12pt]{iopart}

\usepackage{iopams}
\usepackage{graphicx}
\usepackage{amssymb}
\usepackage{cite}
\usepackage{hyperref}
\hypersetup{
 pdfnewwindow=true,
 colorlinks=true,
 linkcolor=blue,
 citecolor=blue,
 filecolor=blue,
 urlcolor=blue
}
\usepackage{color}
\usepackage{psfrag}
\usepackage{soul}
\usepackage{textcomp}
\usepackage{bbm}

\newcommand{\eqref}[1]{(\ref{#1})}
\newcommand{\comments}[1]{}
\newcommand{\bea}{\begin{eqnarray}}
\newcommand{\eea}{\end{eqnarray}}
\newcommand{\ket}[1]{| #1 \rangle}

\newcommand{\bra}[1]{\langle #1 |}

\newcommand{\erg}{\mathcal E}
\newcommand{\av}[1]{\left\langle #1 \right\rangle}

\newcommand{\id}{\mathbbm{I}}

\newcommand{\bsz}{ \bar{\sigma}_0^z}

\begin{document}

\title[Quantum batteries at the verge of a phase transition]{Quantum batteries at the verge of a phase transition}

\author{Felipe Barra$^1$, Karen V. Hovhannisyan$^{2,3}$ and Alberto Imparato$^4$}
\address{$^1 \;$ Departamento de F\'isica, Facultad de Ciencias F\'isicas y Matem\'aticas, Universidad de Chile, Santiago, Chile}
\address{$^2 \;$ Institute of Physics and Astronomy, University of Potsdam, 14476 Potsdam, Germany}
\address{$^3 \;$ The Abdus Salam International Centre for Theoretical Physics (ICTP), Strada Costiera 11, 34151 Trieste, Italy}
\address{$^4 \;$ Department of Physics and Astronomy, Aarhus University, Ny Munkegade 120, 8000 Aarhus, Denmark}
\ead{imparato@phys.au.dk}


\begin{abstract}
Starting from the observation that the reduced state of a system strongly coupled to a bath is, in general, an athermal state, we introduce and study a cyclic battery--charger quantum device that is in thermal equilibrium, or in a ground state, during the charge storing stage. The cycle has four stages: the equilibrium storage stage is interrupted by disconnecting the battery from the charger, then work is extracted from the battery, and then the battery is reconnected with the charger; finally, the system is brought back to equilibrium. At no point during the cycle are the battery--charger correlations artificially erased. We study the case where the battery and charger together comprise a spin-1/2 Ising chain, and show that the main figures of merit---the extracted energy and the thermodynamic efficiency---can be enhanced by operating the cycle close to the quantum phase transition point. When the battery is just a single spin, we find that the output work and efficiency show a scaling behavior at criticality and derive the corresponding critical exponents. Due to always present correlations between the battery and the charger, operations that are equivalent from the perspective of the battery can entail different energetic costs for switching the battery--charger coupling. This happens only when the coupling term does not commute with the battery's bare Hamiltonian, and we use this purely quantum leverage to further optimize the performance of the device.
\end{abstract}


\section{Introduction}

Quantum batteries store and deliver energy to a quantum system coherently. For such a device, energy leaking during the storing phase is a key issue \cite{Barra2019, Liu2019, Hovhannisyan2019, Farina2019, Pirmoradian2019, Santos2019, Gherardini2020, Kamin2020, Hovhannisyan2020, Zhao2021} that is absent if kept at thermodynamic equilibrium \cite{Barra2019, Hovhannisyan2019, Hovhannisyan2020}. This observation motivated analyzing quantum systems in thermodynamic equilibrium as candidates for quantum batteries \cite{Barra2019, Hovhannisyan2020}. In particular, in Ref.~\cite{Hovhannisyan2020}, we showed that a system (the battery), strongly coupled to a bath (the charger), can efficiently store energy, avoiding leakage. An agent can successively deliver such energy to another quantum system after disconnecting the battery from the bath. To have a meaningful definition of efficiency, one has to close the cycle and reconnect the battery to the charger. After that, the total system thermalizes either under the influence of an external bath or, when the charger is large, as a result of internal evolution (in which case only local observables thermalize \cite{Robinson1973, Bach2000, Gogolin2016, Farrelly2017}). The cycle has an energy cost, the ratio of which to the energy delivered by the battery defines the efficiency.

In Ref.~\cite{Hovhannisyan2020}, we considered the case where the battery and the charger become uncorrelated after the battery is disconnected and discharged. In this paper, we lift that restriction: the battery and the charger keep their correlations after the energy extraction process. This opens up new possibilities to optimize battery performance in terms of efficiency, as discussed in the following.

Additionally, we study the regime in which the battery--charger system is at a quantum phase transition point during the charge storage stage, revealing the effect of criticality on the performance of the device. In particular, when the total system is a quantum lattice, we observe an increase in the efficiency near the quantum critical point. Interestingly, second-order phase transitions (quantum or classical) are known to boost thermodynamic performance in a variety of thermal devices \cite{Golubeva2012a, Golubeva2013, Golubeva2014, Imparato2015, Campisi2016, Ma2017, Herpich2018, Herpich2018a, Sune2019a, Abiuso2020, Imparato2021, Puebla_2021}. However, in those instances, it is the working medium that is critical, and the enhancement is related to the increased collectivity of its constituents due to strong and long-range correlations (exhibited by both classical \cite{Chaikin} and quantum \cite{Sachdev2011} critical systems). Whereas in the present case, the working medium (the battery) consists of just one or two spins, and therefore the observed enhancement is of a fundamentally different nature.

\begin{figure}[h]
\center
\psfrag{ }[ct][ct][1.]{ }
\includegraphics[width=8cm]{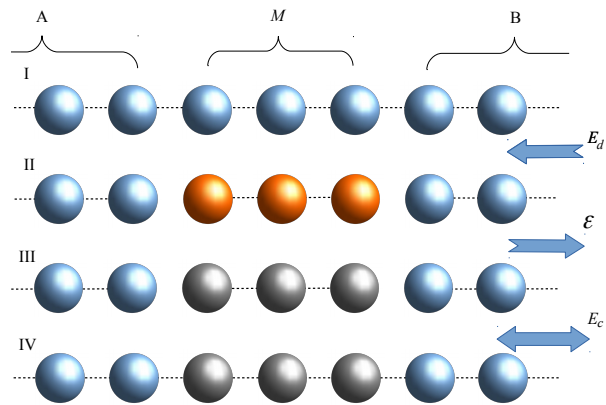}
\caption{Graphic representation of the thermodynamic cycle, made of four strokes. A set of $M$ units (here $M=3$) is disconnected (${\rm I}\to {\rm II}$) from the rest of the chain (represented by the $A$ and $B$ parts). The ergotropy is extracted in ${\rm II}\to {\rm III}$, and finally the exhausted subsystem is reconnected to the chain (${\rm III}\to {\rm IV}$).}
\label{fig:cycle}
\end{figure}

Our workhorse is the 1D transverse-field spin-1/2 Ising chain, whose critical behavior has been fully characterized both in the ground state \cite{Pfeuty1970} and in the thermal state \cite{Barouch1970, Barouch1971, Osborne2002}.
We consider the thermodynamic cycle depicted in figure~\ref{fig:cycle}, where a subset of spins (the battery) is disconnected from the chain initially in the ground or in a thermal state; energy is extracted from it; and the exhausted battery is then reconnected to the rest of the chain playing the role of the charger. We show that the optimal working regime of such a quantum battery occurs on the brink of a phase transition. As figure of merits, we use both the thermodynamic efficiency and the extracted energy, expressed in terms of the battery ergotropy. The ergotropy, defined as the maximum extractable energy from a system in a cyclic unitary process \cite{Allahverdyan2004} is appropriate in our context since we are interested in systems that deliver the energy coherently. Besides focusing on the phase transition, we emphasize the importance of correlation between the subset-of-spins and rest-of-the-chain during the cycle.  
Indeed the presence of strong coupling and battery--charger correlation brings out the fact that locally equivalent operations can have very different global manifestations. We observe that a set of phases of the unitary operator that extracts the battery's ergotropy, which are irrelevant for the (reduced) state of the battery, play a significant role in the reconnecting energy when the battery--charger correlations are taken into account.
This provides us with an additional set of parameters that, as we will see in the following, can be tuned so as to further increase the cycle efficiency.

The paper is organized as follows. In section~\ref{sec.cycle}, we first describe the cycle for implementing a quantum battery, discuss the figures of merits and the role of  local phase manipulation in the energetics of the full system. Then in section~\ref{sec.ising}, we introduce the system, a transverse-field 1D spin-1/2 Ising model, in which the cycle is studied and summarize the statistical properties of the chain in the ground state and the thermal state. Section~\ref{sec:M1}  illustrates our results for a single spin battery in the limit of an infinite chain. Here, we can derive the exact critical exponent characterizing the ergotropy around the critical point. In section~\ref{sec:MT}, we study numerically larger batteries and finite chargers in initial thermal states. We conclude in section~\ref{sec:conclu}.

\section{The working cycle}
\label{sec.cycle}

The working cycle is depicted in Fig.~\ref{fig:cycle}. We do not consider a specific working substance at this stage: the cycle requires a system (the battery) that is strongly coupled to a bath (the charger) and the ability of an agent to connect and disconnect the system from the bath. As illustrated in Fig.~\ref{fig:cycle}, we consider system units and bath units of the same type.
We assume the ability to perform any unitary operation on the battery without affecting the coherence between the battery and the charger, i.e., we treat them as an isolated quantum system as we perform stages I to IV of the cycle.  The closing step IV $\to$ I might involve coupling to some external systems, for instance, a weak coupling to a super-bath or, if the battery--charger system is large, internal evolution causing return to equilibrium \cite{Robinson1973, Bach2000, Gogolin2016, Farrelly2017}. Either way, this step is not relevant for the energetic budget of the agent. In this sense, we assume that the initial thermal or ground state, denoted by $\varrho_{\rm I}$, is a resource given to us.

The available resources and operations we have described above are very similar to those considered in \cite{Hovhannisyan2020} except for the fact that there, in the reconnecting stage III $\to$ IV, the battery was uncorrelated to the bath (charger). This fundamental difference affects the efficiency but not  the ergotropy, as we discuss below. 

The Hamiltonian of the total battery--charger system is
\begin{equation}
H_{\rm tot}=H_S+H_R+H_{\rm int}
\end{equation} 
where $H_S$ and $H_R$ are the Hamiltonians of the battery ``system'' $S$  and of the charger $R$ respectively. The interaction Hamiltonian between $S$ and $R$ is  $H_{\rm int}$. In the following we will use the symbol $\varrho$ to indicate the state of the total system, while the symbol $\rho$ will be used for the reduced state of $S$---the battery---alone.

The cycle consists of the following four strokes:
\begin{itemize}
\item In the stroke I $\to$ II, with the battery and charger in the state $\varrho_{\rm I}$, we instantaneously disconnect $S$ from $R$. The energy cost for the quench reads
\begin{equation}
E_d = - \tr[ H_{\rm int} \varrho_{\rm I}]
\label{Ed:eq}
\end{equation}

Immediately after the quench, at the beginning of stage II, the system $S$ will be in the reduced state
\begin{equation}
\rho_{\rm II} = \tr_{R} [\varrho_{\rm I}].
\label{rhoiis:eq}
\end{equation} 

\item Given $\rho_{\rm II}$ and $H_S$ we can extract the ergotropy $\erg$ from $S$ (stroke  II $\to$ III), taking $\rho_{\rm II}$ to $\rho_{\rm III}$, where
\begin{equation} 
\rho_{\rm III} = U_{\erg} \, \rho_{\rm II} \, U_{\erg}^\dagger
\label{rhoiii:eq}
\end{equation}
is now the exhausted (or passive) state of $S$ and $U_{\erg}$ is a unitary operator that extracts the ergotropy $\erg$ from the system.

Assuming that this step is also instantaneous, from the perspective of the total system, the II $\to$ III transition results in the new state 
\begin{equation} \label{rho:3}
\varrho_{\rm III} = {\mathcal U}_{\erg} \, \varrho_{\rm II} \, {\mathcal U}_{\erg}^\dagger = {\mathcal U}_{\erg} \, \varrho_{\rm I} \, {\mathcal U}_{\erg}^\dagger 
\end{equation}
where the second equality  is due to the fact that $\varrho_{\rm II} = \varrho_{\rm I}$, and the total unitary operator reads
\begin{equation}
{\mathcal U}_{\erg} =U_{\erg} \otimes \id_R.
\end{equation} 
The identity operator acting on $R$ manifests the assumption of a fast ergotropy extraction and represent a simplification. The important assumption is that the full system evolves unitarily and that there is no control or manipulation of the bath $R$. 

The ergotropy thus reads
\begin{equation} 
\erg = \tr[ H_S (\rho_{\rm II} - \rho_{\rm III})] .
\label{ergo:eq}
\end{equation} 

\item In the next stroke III $\to$ IV we suddenly reconnect $S$ to $R$. The energy cost of this operation reads
\begin{equation} 
E_c = \tr[H_{\rm int} \varrho_{\rm III}]=\tr[{\mathcal U}_{\erg}^\dag H_{\rm int}{\mathcal U}_{\erg} \varrho_{\rm II}],
\label{Ec:eq}
\end{equation} 
but the state is unchanged: $\varrho_{\rm IV} = \varrho_{\rm III}$.

\item Lastly, to close the cycle, we may perform the step VI $\to$ I and bring the system back to its initial state $\varrho_{\rm I}$ by, e.g., connecting the full system weakly to a super-bath. In that case, the energy (heat) delivered to the total system will be
\begin{equation}
E_{\rm th} = \tr[(\varrho_{\rm I}-\varrho_{\rm IV}) H_{\rm tot}],
\label{Eth:eq}
\end{equation} 
and it is not a cost for the agent running the cycle. In the case when the total system is left to rethermalize by itself (e.g., when the total system is large and the temperature is $>0$), the energetic cost of that will of course be zero. Although the system will locally appear thermal, self-rethermalization does affect the global state, and therefore the price of zero-energy reset is that the energetics of the cycle will be affected in the long run.
\end{itemize}

In the following, we shall study the maximal work $\erg$ that can be extracted from the battery (the ergotropy) during the cycle, Eq.~(\ref{ergo:eq}), and the cycle efficiency \cite{Hovhannisyan2020}, as given by
\begin{equation}
\eta=\frac{\erg}{E_c+E_d}.
\label{eta:def}
\end{equation} 
We notice that in this expression both $\erg$ and $E_c$ are determined by the the ergotropy-extracting unitary operator $U_{\erg}$ appearing in Eq.~(\ref{rhoiii:eq}).

\subsection{The unitary $U_{\erg}$}
\label{subsecUgen}

The ergotropy-extracting operator $U_{\erg}$, appearing in Eq.~(\ref{rhoiii:eq}), is a unitary that achieves the minimization of the final energy, defining \cite{Allahverdyan2004} the ergotropy:
\bea \nonumber
\erg=\Tr[H_S\rho_{\rm II}]-{\rm min}_U \Tr[H_SU\rho_{\rm II} U^\dag].
\eea
For $U_\erg$, an explicit expression can be found in terms of the normalized eigenvectors of $\rho_{\rm II}$ and $H_S$ \cite{Allahverdyan2004}. Consider the spectral decompositions of $\rho_{\rm II}$ and $H_S$
\bea
\rho_{\rm II} &=& \sum_{\alpha = 1}^{2^M} r_\alpha^\downarrow \ket{r_\alpha^\downarrow} \bra{r_\alpha^\downarrow},
\\
H_S &=& \sum_{\alpha = 1}^{2^M} \epsilon_\alpha^\uparrow \ket{\epsilon_\alpha^\uparrow} \bra{\epsilon_\alpha^\uparrow},
\eea
where $\downarrow$ and $\uparrow$ indicate that the eigenvalues are ordered, respectively, decreasingly and increasingly. We can thus write $U_{\erg}$ in Eq.~\eqref{rhoiii:eq} as
\begin{equation}
U_{\erg}[\, \vec{\theta} \, ] := \sum_\alpha e^{i \theta_\alpha} \ket{\epsilon_\alpha^\uparrow} \bra{r_\alpha^\downarrow},
\label{ugeneral}
\end{equation} 
where $\vec{\theta} = \{ \theta_\alpha \}_\alpha$ is a $2^M$-tuple of arbitrary real numbers, manifesting the arbitrariness of the normalized eigenvectors $\ket{\epsilon_\alpha^\uparrow}$ and $\ket{r_\alpha^\downarrow}$. Note that one of these numbers determines a global phase and thus we can reduce their number to $2^M-1$.

Usually, these phases are omitted (i.e. one takes $\theta_\alpha=0, \,\forall \alpha$) since $\rho_{\rm III} = U_{\erg} \rho_{\rm II} U_{\erg}^\dag = \sum_\alpha r_\alpha^\downarrow \ket{\epsilon_\alpha^\uparrow} \bra{\epsilon_\alpha^\uparrow}$ and thus neither the final passive state $\rho_{\rm III}$ nor the ergotropy $\erg$ depend on them. We note here, for later convenience, that $\rho_{\rm III}$ is diagonal in the energy basis $\left\{\ket{\epsilon_\alpha^\uparrow}\right\}$ and the population decreases as the energy increases; these are the so-called passive states \cite{Pusz1978, Lenard1978}, characterized by the property that no energy can be extracted from them through a cyclic unitary process.

Thus, the operator $U_{\erg}$ is not unique even if the spectra of $\rho_{\rm II}$ and $H_S$ are non-degenerate. However, while this freedom is irrelevant for any observable property of $S$, the choice of $\vec{\theta}$ will affect the global state $\varrho_{\rm III}$ [cf. Eq.~\eqref{rho:3}]. As a result, it will affect $E_c$; see Eq.~\eqref{Ec:eq}. In order for this effect to occur, it is essential that the coherence, manifested by the unitary evolution of the full chain, and the correlation between the battery and the charger, are maintained during the steps I $\to$ IV.

Indeed, had we considered a different setup such that after the stroke II $\to$ III the correlations between $S$ and $R$ were lost, we would have obtained another state $\varrho_{\rm III}' = \rho_{\rm III} \otimes \omega_R$ at the end of the stroke III. The state of the charger $\omega_R$ before the reconnection stroke (III $\to$ IV) could be the reduced state of the charger after stroke II $\to$ III or another ``fresh'' charger state, as in Ref.~\cite{Hovhannisyan2020}. The reconnecting energy $\Tr[H_{\rm int} \rho_{\rm III}\otimes\omega_R]$ would then be independent of $\vec{\theta}$. 

To summarize this section, when the coherence and correlations are preserved during the evolution of the total system during the first three strokes, the phases of the eigenstates of $\rho_{\rm III}$ and $H_S$ influence the connecting energy $E_c(\vec{\theta})$ and thus the efficiency of the cycle $\eta$. Such phases can in principle be manipulated by the agent extracting the ergotropy, and in the following we investigate the effect of these phases on the cycle and use them as free parameters to optimize the performance of a battery--charger system made of 1/2 quantum spins.

\subsection{Remarks on the thermodynamics of the cycle}

As a result of the first three strokes, the state of full system evolves unitarily: $\varrho_{\rm I} \to \varrho_{\rm IV}={\mathcal U}_{\erg}\varrho_{\rm I}{\mathcal U}_{\erg}^\dag$; whereas the total Hamiltonian is changed cyclically---$H_{\rm tot}^{\rm (IV)} = H_{\rm tot}^{\rm (I)}$. Since the initial state of the total system is passive due to the fact that it is either a Gibbs state or a ground state \cite{Pusz1978, Lenard1978}, this can only increase its average energy:
\bea
\Tr[H_{\rm tot}(\varrho_{\rm IV}-\varrho_{\rm I})]\geq 0.
\label{secondlaw:eq}
\eea
In other words, one can perform only positive work on the total system.

Therefore, since the work performed on the total system during the first three cycles is $E_d - \erg + E_c$, we have that $E_d + E_c \geq \erg$. Since, by definition, $\erg \geq 0$, the definition of efficiency in Eq.~\eqref{eta:def} is indeed meaningful and, moreover, $\eta \leq 1$.

Finally, if the closing step IV $\to$ I is achieved by coupling the system weakly to a bath, the dissipation (entropy production) for the cycle will be $-E_{\rm th}/T$, where $T$ is the temperature of the bath. In view of Eqs.~\eqref{Eth:eq} and \eqref{secondlaw:eq}, $-E_{\rm th}/T\geq 0$, as one would expect from the second law.

In our previous work \cite{Hovhannisyan2020}, additional sources of dissipation included the change $\varrho_{\rm III}\to\rho_{\rm III}\otimes\omega_R$ modeling the loss of correlation between the battery and the charger, plus the ``refreshing'' of the charger; those are absent in the present setup.

\section{The working substance: transverse spin-1/2 Ising chain}
\label{sec.ising}

We introduce now the specific working substance we use to study the cycle depicted in Fig.~\ref{fig:cycle}---the transverse spin-1/2 quantum Ising chain described by the Hamiltonian
\begin{equation}
H_{\rm tot} = -(1-f)\sum_{i=0}^{N-1} \sigma^x_i \sigma^x_{i+1} - f \sum_{i=0}^{N-1} \sigma^z_i,
\label{HN:def}
\end{equation} 
with periodic boundary conditions (PBC) $\sigma^{\alpha}_{N}=\sigma^{\alpha}_{0}$, the latter ensuring translation symmetry and reflection symmetry around any site. Some (completely different) thermodynamic aspects of subsystems of the quantum Ising chain were studied in Ref.~\cite{Campisi2010}.

The battery $S$ consists of $M$ consecutive nodes of the chain, and the charger $R$ consists of the remaining nodes; see Fig.~\ref{fig:cycle} for an illustration. The interaction Hamiltonian between $S$ and $R$ is
\bea
H_{\rm int}= 
- (1-f)( \sigma^x_{i-1}  \sigma^x_{i} + \sigma^x_{i+M-1}  \sigma^x_{i+M}) \label{Hint},
\eea
with $i$ arbitrary given the PBC, and 
where $H_S$ and  $H_R$ are the bare Hamiltonian of $S$ and $R$, respectively:
\bea
H_X &=& -(1-f) \sum_{\{j\} \cup \{j+1\} \subseteq X}  \, \sigma^x_j  \sigma^x_{j+1} - f \sum_{j \in X} \sigma^z_j, 
\eea
with $X$ being either $S$ or $R$.

When $N\to \infty$, the system described by the Hamiltonian (\ref{HN:def}) presents a quantum phase transition at $f_c=1/2$. Using some known results about the quantum Ising model \cite{Pfeuty1970, Barouch1970, Barouch1971}, we can study our battery analytically for $M=1$ and partially for $M=2$. For general $M$ and $N$, we will have to analyze the problem numerically.

\subsection{Transverse Ising chain in the ground state}

Let us review some of the well-known properties of the ground state of the transverse Ising chain \cite{Pfeuty1970, Barouch1970, Barouch1971, Osborne2002}. As mentioned above, the system presents a quantum phase transition at $f_c = 1/2$: the ground state $\ket{0}$ is not degenerate for $f>1/2$, but becomes doubly degenerate $\ket{0^\pm}$ for $f<1/2$. 
The longitudinal magnetization $\av{\sigma_i^x}$ changes from a vanishing value for $f \geq 1/2$ to a positive  
or negative value for $f<1/2$, depending on the ``branch'' of the ground state. Without loss of generality, in the discussion that follows, we choose the system to be in the eigenstate $\ket{0^+}$ for $f<1/2$, so that $\av{\sigma_i^x}\geq 0$. 
Setting $\lambda=(1-f)/f$, the following formulas for the longitudinal magnetization (the ``order parameter'') hold:
\bea
\av{\sigma_i^x}=
\left\{
\begin{array}{cc}
(1-\lambda^{-2})^{\beta}, & f < 1/2 \; (\lambda >1) \\
0, & f \geq 1/2 \; (\lambda \leq 1)
\end{array} \right.
\label{sigmax}
\eea
with the critical exponent $\beta=1/8$.

The transverse magnetization reads 
\bea
\av{\sigma_i^y}&=&0,\\
\av{\sigma_i^z}&=&
\frac{1}{\pi}\int_0^\pi d\phi\frac{1+\lambda \cos\phi}{\sqrt{1+\lambda^2+2\lambda \cos\phi}}.
\label{sigmaz}
\eea 
Note that $\av{\sigma_i^z}$ changes smoothly at the transition and is positive for positive $f$.

The two-site correlators read \cite{Pfeuty1970}: 
\bea
\label{corrxy}
\av{\sigma_i^x\sigma_{i+1}^y}&=& 0, \\
\av{\sigma_i^y\sigma_{i+1}^z}&=& 0, \\
\label{corrxx}
\av{\sigma_i^x\sigma_{i+1}^x}&=&\frac{1}{\pi}\int_0^\pi d\phi\frac{\cos\phi+\lambda}{\sqrt{1+\lambda^2+2\lambda\cos\phi}}, \\
\av{\sigma_i^y\sigma_{i+1}^y}&=& \frac{1}{\pi}\int_0^\pi d\phi\frac{\cos\phi+\lambda\cos 2\phi}{\sqrt{1+\lambda^2+2\lambda\cos\phi}}, \\
\av{\sigma_i^z\sigma_{i+1}^z}&=&\av{\sigma_i^z}^2-\av{\sigma_i^x\sigma_{i+1}^x}-\av{\sigma_i^y\sigma_{i+1}^y}.
\label{corr0}
\eea
An analytic expression for $\av{\sigma_i^z\sigma_{i+1}^x}$ has not been found as after the Jordan--Wigner transformation the correlation operator still contains non local terms \cite{Pfeuty1970, Osborne2002}. 

As we will see, $\av{\sigma_i^z\sigma_{i+1}^x}$ determines the reconnecting energy $E_c$ (\ref{Ec:eq}) and thus 
we resort to two different numerical approaches to evaluate it. We first diagonalize directly the Hamiltonian (\ref{HN:def}), for a finite value $N$ of spins, and then with a density matrix renormalization group (DMRG) algorithm. We anticipate that the results are not significantly different in the region of interest, besides a moderate discrepancy for $f$ close to $f_c$; see \ref{num:app} for further details on the numerical methods. 

When the spin chain is in the thermal state $\sim e^{-H_{\rm tot}/k_BT}$, similar expressions can be obtained for the average magnetization and correlations \cite{Barouch1970}; we list them in \ref{App:ChainfiniteT}.

\section{Single spin battery $(M=1)$ in the ground state}
\label{sec:M1}

Of particular interest is the case where only one spin is disconnected from an infinite  chain ($M=1,N=\infty$). Besides being of pedagogical relevance, all but one of the thermodynamic quantities of the single spin battery can be expressed in analytic form as discussed below.

We assume that at the beginning of the cycle the spin chain is in the ground state $\varrho_{\rm I}=\ket{0^+}\bra{0^+}$. 

Since the chain is translation-invariant, the choice of the battery site is arbitrary; for definiteness, we choose it to be the zeroth site. In the stroke I$\to$II, the battery spin is instantaneously disconnected with a work cost \bea
E_d=2(1-f)\av{\sigma_0^x\sigma_1^x},
\label{EdM1:eq}
\eea
where  $\av{\dots} $ is the expectation value calculated over the initial state $\varrho_{\rm I}$. Note that we have exploited the translation-invariance of the chain for the nearest-neighbor correlators. Moreover, it is easy to see that Eq.~\eqref{EdM1:eq} is independent of the number of battery sites $M$.

The reduced state of the single spin (\ref{rhoiis:eq}) after the disconnection quench thus reads
\bea
\rho_{\rm II}=\frac{1}{2}(\id_2+\av{\sigma_0^x}\sigma_0^x+\av{\sigma_0^z}\sigma_0^z)=\frac{1}{2}(\id_2+\bf{a \cdot} {\boldsymbol\sigma}_0)
\label{rho0:eq}
\eea
represented by the vector ${\bf a}=(\av{\sigma_0^x},0,\av{\sigma_0^z})$ in the Bloch sphere in terms of \eqref{sigmax} and \eqref{sigmaz}. 

In the stroke II$\to$III, a unitary operation extracts the ergotropy of the state $\rho_{\rm II}$. Since the Hamiltonian $H_S=-f\sigma_0^z$, and the exhausted (or passive) state of the disconnected spin commute, the Bloch vector of the exhausted state must point in the $\bf{z}$ direction.
Unitary transformations on the spin corresponds to rotations of the Bloch vector ${\bf a}$. Thus, the unitary that extracts the ergotropy rotates the Bloch vector from ${\bf a}$ to $\bar{{\bf a}}=(0,0,\bar{\sigma_0}^z)$, where
\begin{eqnarray}
 \bsz&=&\sqrt{\av{\sigma_0^x}^2+\av{\sigma_0^z}^2}\label{sz1}.
\end{eqnarray} 
I.e., the passive state (\ref{rhoiii:eq}) is
\begin{eqnarray}
\rho_{\rm III}&=&\frac{1}{2}(\id+\bsz\sigma_0^z).
\label{rho01}
\end{eqnarray} 
It is simple to see that the Bloch vector ${\bf a}$ was rotated an angle $2\alpha$ with respect to the $\hat{\bf y}$ axis, where 
\bea
\label{Eq19}
\sin 2\alpha=\frac{\av{\sigma_0^x}}{\bsz}\\
\cos 2\alpha=\frac{\av{\sigma_0^z}}{\bsz}.
\label{Eq20}
\eea
The ergotropy (\ref{ergo:eq}) of the single spin will thus be given by
\bea
\erg=\frac{f}{2}\Tr[\sigma_0^z(\bsz\sigma_0^z-\av{\sigma_0^x}\sigma_0^x-\av{\sigma_0^z}\sigma_0^z)] =f(\bsz-\av{\sigma_0^z}).
\label{ergo:theo}
\eea

Comparing the last expression with eq.~(\ref{sz1}), we conclude that we need $\av{\sigma_0^x } \neq 0$, i.e., $f<1/2$ for the ergotropy to be non vanishing. In other words, the battery is charged only in the ordered phase.

The magnetization along $x$, given by Eq.~(\ref{sigmax}), 
grows smoothly from zero at $f_c=1/2$ as $f$ decreases. 
By expanding (\ref{ergo:theo}) to the leading order, we can thus obtain the critical behaviour of the ergotropy
\begin{equation}
\erg\sim (f_c-f)^{1/4}+O ((f_c-f)^{1/2}).
\label{erg:crit}
\end{equation} 
Thus we find that the ergotropy critical exponent is $2\beta$, where $\beta$ is the critical exponent for the order parameter $\av{\sigma_i^x}$: the two  critical exponents are not independent, akin to the scaling relations in critical systems \cite{Chaikin}. Eq.~(\ref{ergo:theo}), together with Eq.~(\ref{erg:crit}), represents the first relevant result in this section.

\begin{figure}[h]
\center
\includegraphics[width=8cm]{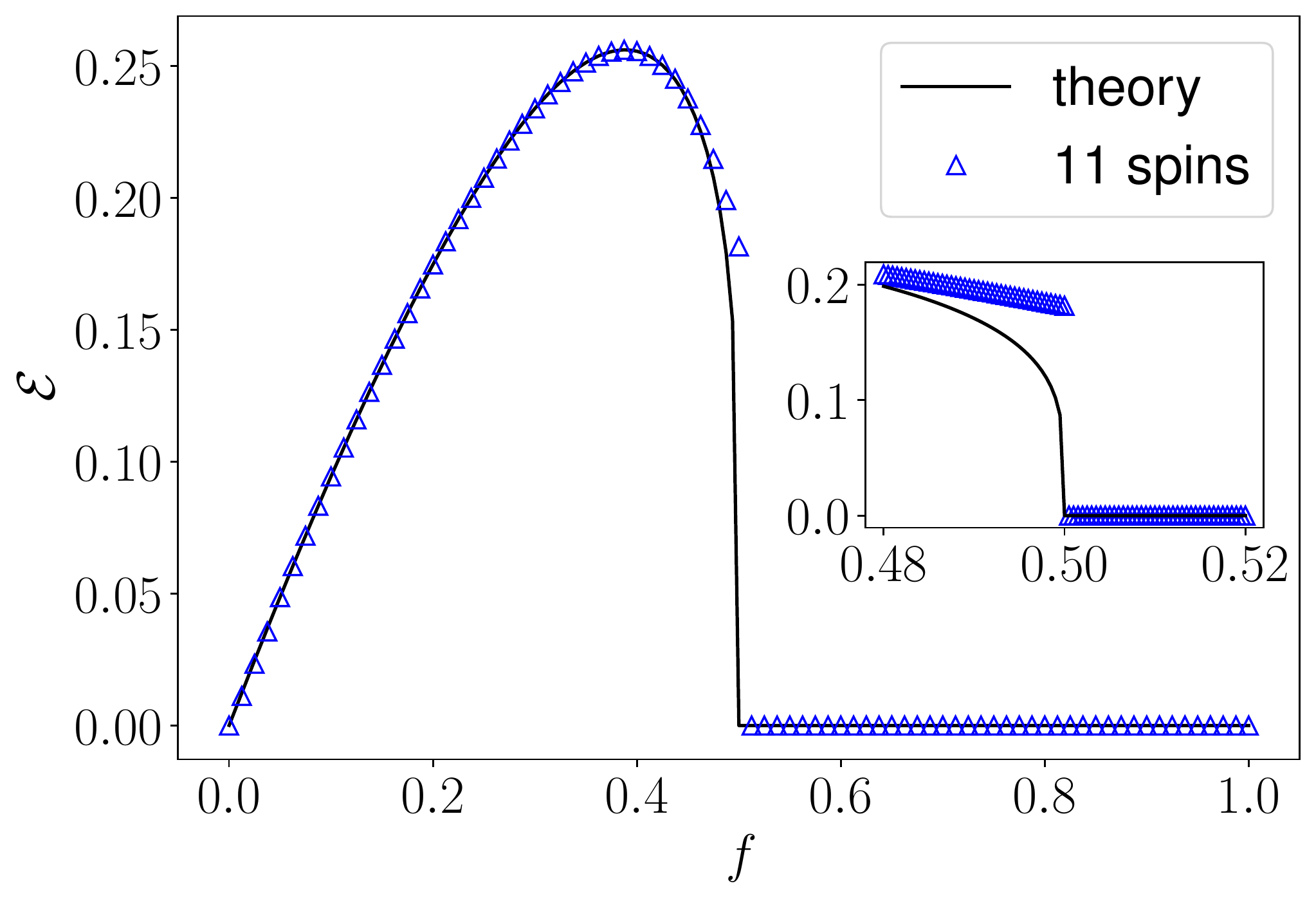}
\caption{Full line: Ergotropy of the single spin battery ($M=1$) in the ground state as a function of $f$, as given by Eq.~(\ref{ergo:theo}). Points: numerical approximation of the ergotropy as obtained by diagonalising the Hamiltonian (\ref{HN:def}) with $N=11$ spins. Inset: zoom of the plot in the critical region.}
\label{fig:ergo}
\end{figure}

Finally we note that the ergotropy as a function of $f$ vanishes at $f=0$ and for $f>1/2$, and noticing that $\erg/f=\bsz-\av{\sigma_0^z}$ decreases monotonically from its maximum $\erg/f=1$ at $f=0$ to $\erg/f=0$ at $f=1/2$ we conclude that $\erg$ must have a single maximum in the interval $0<f<1/2$. 
This is confirmed by inspection of Fig.~\ref{fig:ergo} where  a plot of the ergotropy is shown.

To compute the ergotropy (\ref{ergo:theo}) we did not write explicitly the unitary operator $U_{\erg}$ introduced in Eq.~(\ref{rhoiii:eq}). We noticed that it corresponds to a rotation around the $\hat{\bf y}$ axis bringing the ${\bf a}$ vector to the $\hat{\bf z}$ direction. On the Hilbert space of the single spin the corresponding unitary operator is $e^{i\alpha\sigma_0^y}$ with $2\alpha$, see Eqs.\eqref{Eq19} and \eqref{Eq20}, the angle of rotation in the sphere \cite{Haroche}. Once the Bloch vector points towards the $\hat{\bf z}$ direction, an arbitrary rotation around that axes leaves the state invariant, i.e., given $\alpha$, for any $\theta$, the unitary operator
\bea
\label{Using:theta}
U_{\erg}(\theta)=e^{i\theta\sigma_0^z}e^{i\alpha\sigma_0^y}
\eea
extracts the ergotropy. 
In~\ref{appendixUM1}, we derive the same expression for  $U_{\erg}(\theta)$ starting from Eq.~\eqref{ugeneral} with  $M=1$

 As discussed in section \ref{subsecUgen}, neither the value of the ergotropy as given by Eq.~(\ref{ergo:eq}) nor $\rho_{\rm III}$ in Eq.~(\ref{rhoiii:eq}) depend on  the phase $\theta$, while $\varrho_{\rm III}$ and $E_c$ do.

 We are now in the conditions to calculate the reconnecting work $E_c$, Eq.\eqref{Ec:eq}, after the ergotropy extraction (step III $\to$ IV in Fig.~\ref{fig:cycle}).
Such a quantity for the $M=1$ case reads: 
\begin{eqnarray}
E^{(1)}_c
&=& 2 (1-f) \cos 2 \theta (\sin 2 \alpha \av{\sigma^x_0\sigma^z_{1}}- \cos2\alpha\av{\sigma^x_0\sigma^x_{1}} ),
\label{Wc:def}
\end{eqnarray} 
where we have used the fact that the correlations do not depend on the specific site, that $\av{\sigma^x_0\sigma^y_{1}}=0$, see Eq.~(\ref{corrxy}), and that the equality $\av{\sigma^x_i\sigma^z_{i+1}}=\av{\sigma^z_i\sigma^x_{i+1}}$ holds due to the inversion and translation symmetry. 

By using eqs.~(\ref{Eq19})--(\ref{Eq20}), or the more explicit expression eqs.~(\ref{Eq19b})--(\ref{Eq20b}), the connecting energy  Eq.~(\ref{Wc:def})
can be written as
\begin{equation}
E^{(1)}_c=2(1-f)  \frac{\cos 2 \theta}{\bsz}(\av{\sigma_0^x} \av{\sigma^x_0\sigma^z_{1}}-\av{\sigma_0^z} \av{\sigma^x_0\sigma^x_{1}}).
\label{Wc:defp}
\end{equation} 

\begin{figure}[h]
\center
\psfrag{ }[ct][ct][1.]{ }
\includegraphics[width=7.5cm]{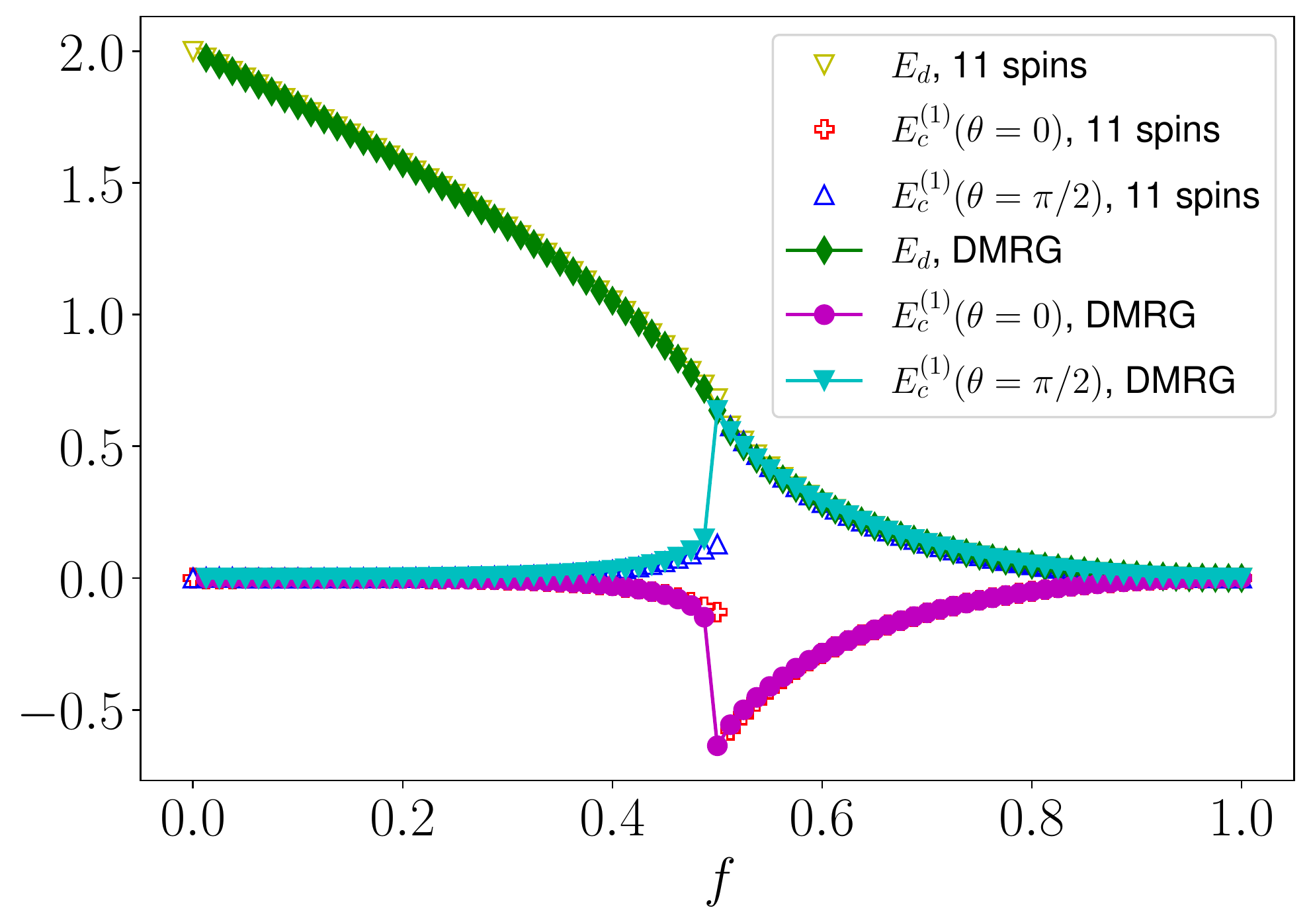}
\includegraphics[width=7.5cm]{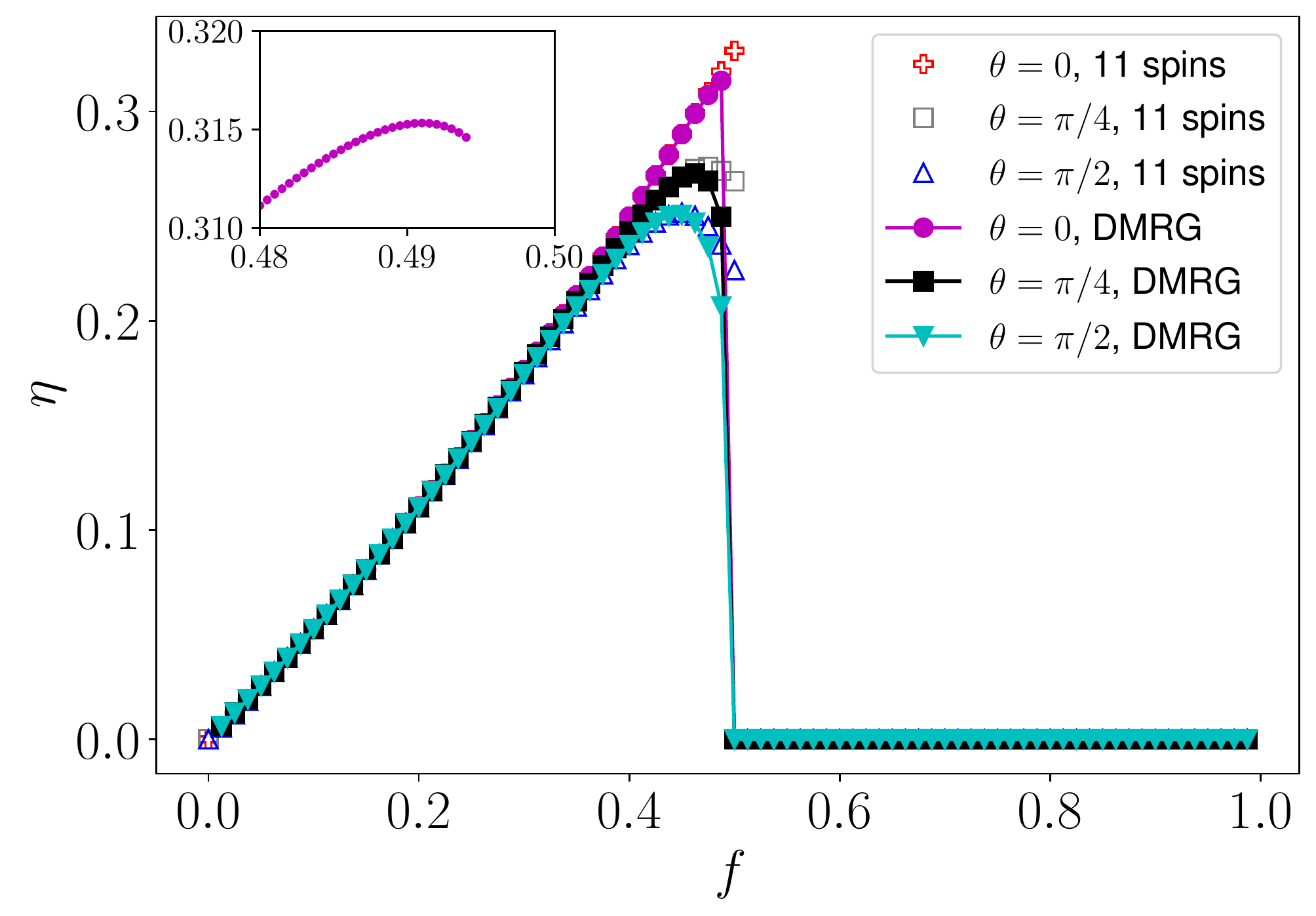}
\caption{Left: disconnecting  and connecting energies as functions of $f$, for the single spin battery ($M=1)$ in the ground state of (\ref{HN:def}), as given by Eq.~(\ref{Ed:eq}) and (\ref{Wc:defp}), respectively. Right:  Efficiency of the single spin battery, $\eta=\erg/(E_d+E_c)$ with $\erg$ form Eq.~(\ref{ergo:theo}) and $E_d+E_c$ from Eq.~(\ref{eimput}). In both panels $\av{\sigma^x_i\sigma^z_{i+1}}$ and thus $E_c(\theta)$   are obtained numerically with the DMRG algorithm (full symbols) or by direct diagonalization of the Hamiltonian (\ref{HN:def}) (empty symbols), for a finite system with $N=11$ spins. Inset: zoom of the plot with $\theta=0$ in the critical region.}
\label{best:DMRG}
\end{figure}

Eq.~(\ref{Wc:defp}) is the second relevant result of this section:
we find that while the other two energies involved in the cycle, namely the disconnecting energy $E_d$ and the ergotropy $\erg$, are independent of the arbitrary phase $\theta$ appearing in Eq.\eqref{Using:theta}, the reconnecting energy does depend on this phase.
In particular, one can tune it such as to minimize $E^{(1)}_c$, and by noticing that 
$(\av{\sigma_0^x} \av{\sigma^x_0\sigma^z_{1}}-\av{\sigma_0^z} \av{\sigma^x_0\sigma^x_{1}})\le0$
 (numerically checked, data not shown), we conclude that $\theta=0$ corresponds to the minimal value of $E_c$ for any $f$.  The connecting and disconnecting energies are plotted in the left panel of Fig.~\ref{best:DMRG} as functions of $f$ and for different values of the phase $\theta$: the effect of the phase on $E_c$ is clearly visible in the figure.

Having derived the expression for the cycle output energy, the ergotropy [Eq.~(\ref{ergo:theo})], and the input energy 
\bea
E_d+E_c=2(1-f)\left\{\frac{\cos 2 \theta}{\bsz}\av{\sigma_0^x} \av{\sigma^x_0\sigma^z_{1}}+\left(1-\frac{\av{\sigma_0^z}\cos 2 \theta}{\bsz}\right) \av{\sigma^x_0\sigma^x_{1}}\right\}
\label{eimput}
\eea 
[Eq.~(\ref{EdM1:eq}) and Eq.~(\ref{Wc:defp})], we can proceed to study the efficiency (\ref{eta:def}) of the cycle for a single spin ($M=1$)
which is maximized for $\theta=0$.
This behaviour is confirmed by inspection of the right panel in Fig.~\ref{best:DMRG}. We also notice that the maximum of the efficiency is achieved for values of $f$  just below the critical value and decreases abruptly as it approaches it.


One can outline an analysis of the scaling behaviour of the efficiency $\eta=\erg/(E_d+E_c)$ near the critical point.
We notice that for $f\to f_c$, both $\av{\sigma_0^z}$ and  $\av{\sigma_0^x\sigma_1^x}$ (eqs.~(\ref{sigmaz}) and (\ref{corrxx}) respectively) go to $2/\pi$ and thus
$E_c+E_d\to 4\sin^2\theta/\pi$ if $\theta \neq k \pi$ with $k$ an integer, as follows from Eq.~\eqref{eimput}. Therefore, the efficiency exhibits the same critical scaling as the ergotropy [see Eq.~\eqref{erg:crit}]:
\begin{equation}
\eta^{(1)}\sim \av{\sigma_0^x}^2 \simeq (f_c-f)^{1/4}, \quad(\theta \neq k \pi),
\end{equation} 
with exponent $2\beta$.

For $\theta=k \pi$, both $\erg\to 0$ and $E_c+E_d\to 0$ as $f\to f_c$
and one finds that, to the left of the critical point ($f\lesssim f_c$), 
\begin{equation}
\eta^{(1)}\sim \frac{\av{\sigma_0^x}}{\av{\sigma^x_0\sigma_{1}^z}}, \quad(\theta = k \pi).
\label{eta1:1}
\end{equation} 
Therefore it is not possible to derive a critical exponent for the efficiency when $\theta = k \pi$,  because the expression for $\av{\sigma_i^x\sigma_{i+1}^z}$, as discussed above, is not available.
However, the numerical results reported in \ref{num:app} clearly show 
that the correlation  $\av{\sigma_i^x\sigma_{i+1}^z}$ vanishes for $f>f_c$, similarly to  $\av{\sigma_i^x}$. Making the physically reasonable assumption that $\av{\sigma_i^x\sigma_{i+1}^z}$ goes to zero continuously as $f\to f_c^-$ with a scaling $\av{\sigma_i^x\sigma_{i+1}^z}\sim(f_c-f)^\delta$, for thermodynamic consistency of Eq.~(\ref{eta1:1}) the inequality $\delta \le  \beta$ must hold.
Thus for $\theta =k \pi$ the overall critical exponent for $\eta^{(1)}$ is $\beta-\delta<\beta$, which explain the abrupt decrement in the curve for $\theta=0$ as  $f\to f_c^-$ in figure \ref{best:DMRG}.
The numerical results in \ref{num:app} shows that  $\av{\sigma_i^x\sigma_{i+1}^z}$ is well fitted with $\delta=\beta/2$ as $f\to f_c^-$.
Thus we find a scaling law for the critical exponent of the efficiency too: its value is determined by a combination of the critical exponents of the order parameter and of the correlations, and is close to $\beta/2$.

\section{Case $T>0$}
\label{sec:MT}

Given that in a thermal state with $T>0$,  the spin chain (\ref{HN:def}) is always in the paramagnetic phase $\av{\sigma_i^x}=0$ \cite{Osborne2002}, Eq.~\eqref{ergo:theo} implies that no ergotropy can be extracted when a single spin is disconnected ($M=1$) from a thermal state. Therefore, in order to study the thermodynamic properties of the cycle in Fig.~\ref{fig:cycle} at finite temperature, we will consider here the case where $M\geq 2$ spins are disconnected from the chain. We present   a few results for $M=2$ in \ref{appendixM2}.
Obtaining the reduced state (\ref{rhoiis:eq}) and the exhausted state (\ref{rhoiii:eq}) becomes a daunting task as $M$ increases.  
Therefore in the following we will resort on the numerical analysis to obtain the thermodynamic quantities of interest. In the previous section, we have seen that direct diagonalization of the Hamiltonian (\ref{HN:def}), or the DMRG algorithm, give results very close to the exact ones when available.



The initial state of the cycle in Fig.~\ref{fig:cycle} is now $\varrho_{\rm I}=\exp(- H_{\rm tot}/k_B T)/Z_{\rm tot}$.  To compute the ergotropy and the post-ergotropy state $\varrho_{\rm III}$  in Eq.(\ref{rho:3}), one must chooses the phases in $U_{\erg}$ introduced in Eq.(\ref{rhoiii:eq}). 
For the following results, we chose them to minimize the reconnection energy. Thus, we proceed as follows. 

Let us introduce
\bea
\mathcal{U}_\alpha = \ket{\epsilon_\alpha^\uparrow} \bra{r_\alpha^\downarrow} \otimes \id_R,
\label{Ucall}
\eea
so that ${\mathcal U}_{\erg} = \sum_\alpha e^{i \theta_\alpha} {\mathcal U}_\alpha$, and rewrite Eq.~\eqref{Ec:eq} accordingly:
\bea
E_c[\, \vec{\theta} \,] = \sum_{\alpha, \gamma} e^{i(\theta_\alpha - \theta_\gamma)} \tr [H_{\rm int} \mathcal{U}_\alpha \varrho_{\rm I} \mathcal{U}_\gamma^\dagger].
\eea
Further introducing
\bea
A_{\alpha, \gamma} := \big\vert \tr [H_{\rm int} \mathcal{U}_\alpha \varrho_{\rm I} \mathcal{U}_\gamma^\dagger] \, \big\vert \qquad \mathrm{and} \qquad \phi_{\alpha, \gamma} := \arg \tr [H_{\rm int} \mathcal{U}_\alpha \varrho_{\rm I} \mathcal{U}_\gamma^\dagger],
\label{Aag}
\eea
which are quantities that do not depend on $\vec{\theta}$, and noting that $\phi_{\gamma, \alpha} = - \phi_{\alpha, \gamma}$, we finally obtain
\bea
E_c[\, \vec{\theta} \,] = \sum_\alpha A_{\alpha, \alpha} + 2 \sum_{\alpha < \gamma} A_{\alpha, \gamma} \cos (\theta_\alpha - \theta_\gamma + \phi_{\alpha, \gamma}).
\eea

The problem of minimization of $E_c$ can thus be formulated as
\bea
E_c^{\min} = \sum_\alpha A_{\alpha, \alpha} + 2 \; \min_{\vec{\theta} \in [0, 2 \pi) \times \cdots \times [0, 2 \pi)} \; \sum_{\alpha < \gamma} A_{\alpha, \gamma} \cos (\theta_\alpha - \theta_\gamma + \phi_{\alpha, \gamma}).
\label{eq:ecminta}
\eea
We immediately see that, when $[H_{\mathrm{int}}, H_S \otimes \id_R] = 0$ and $H_S$ has a nondegenerate spectrum, $A_{\alpha, \gamma} = 0$ whenever $\alpha \neq \gamma$, meaning that all that $\vec{\theta}$-dependent terms vanish, and hence there is no room for optimizing $E_c$. Indeed, plugging Eq.~\eqref{Ucall} into Eq.~\eqref{Aag} and keeping in mind that, by definition, $H_S \ket{\epsilon_\alpha^\uparrow} = \epsilon_\alpha^\uparrow \ket{\epsilon_\alpha^\uparrow}$, we can write
\bea \nonumber
\left\vert \epsilon_\alpha^\uparrow \right\vert A_{\alpha, \gamma} &=& \left\vert \tr \big[ \id_R \otimes \ket{r_\gamma^\downarrow} \bra{\epsilon_\gamma^\uparrow} H_{\mathrm{int}} (H_S \ket{\epsilon_\alpha^\uparrow}) \bra{r_\alpha^\downarrow} \otimes \id_R \, \varrho_{\mathrm{I}} \big] \right\vert
\\ \nonumber
&=& \left\vert \tr \big[ \id_R \otimes \ket{r_\gamma^\downarrow} \bra{\epsilon_\gamma^\uparrow} (H_S \otimes \id_R) H_{\mathrm{int}} \ket{\epsilon_\alpha^\uparrow} \bra{r_\alpha^\downarrow} \otimes \id_R \, \varrho_{\mathrm{I}} \big] \right\vert
\\ \nonumber
&=& \left\vert \epsilon_\gamma^\uparrow \, \tr \big[ \id_R \otimes \ket{r_\gamma^\downarrow} \bra{\epsilon_\gamma^\uparrow} H_{\mathrm{int}} \ket{\epsilon_\alpha^\uparrow} \bra{r_\alpha^\downarrow} \otimes \id_R \, \varrho_{\mathrm{I}} \big] \right\vert
\\ \nonumber
&=& \left\vert \epsilon_\gamma^\uparrow \right\vert A_{\alpha, \gamma},
\eea
from which, in view of the assumed nondegeneracy of the spectrum of $H_S$, the statement follows immediately.

\begin{figure}[h]
\center
\includegraphics[width=6cm]{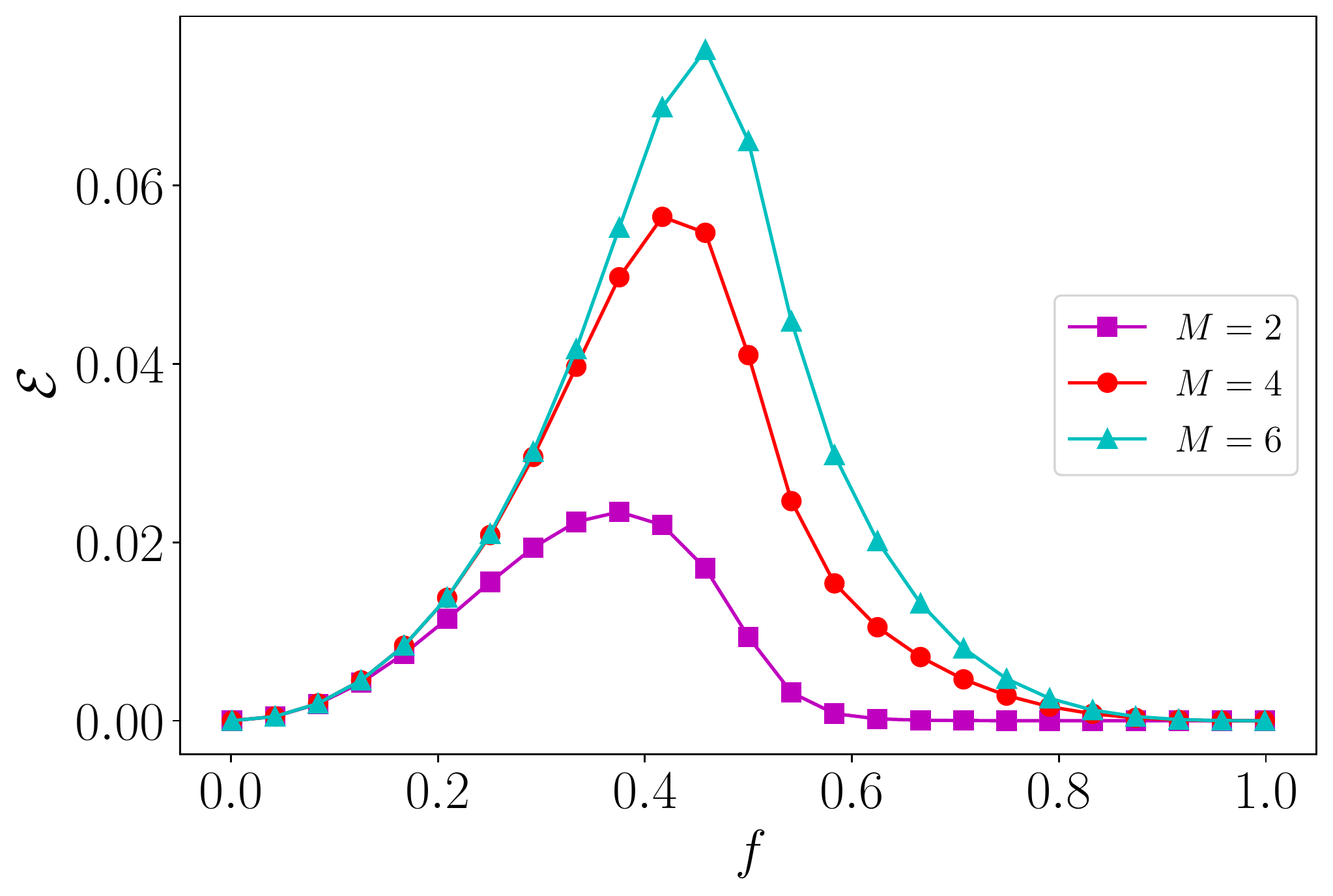}
\includegraphics[width=6cm]{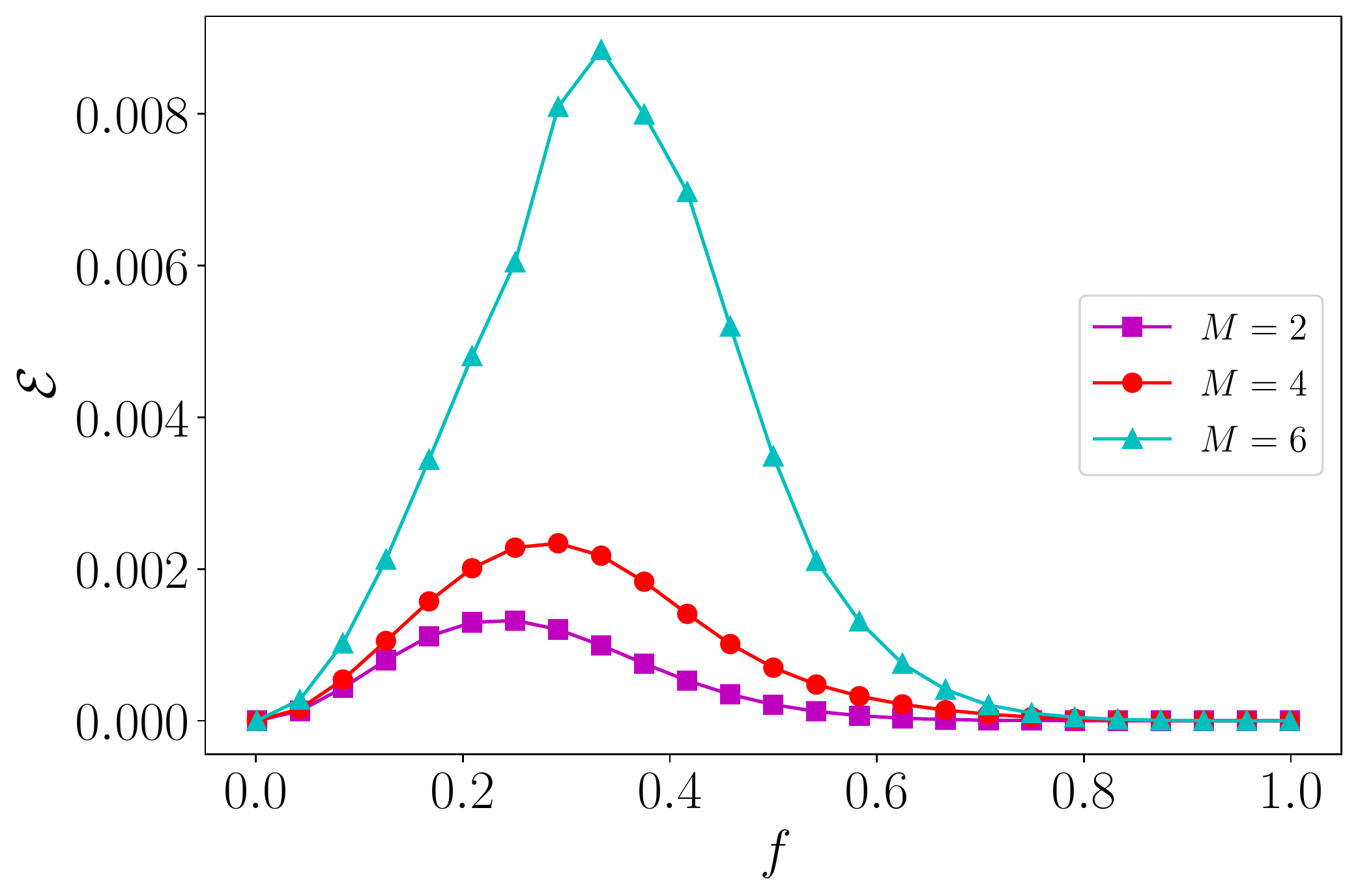}
\includegraphics[width=6cm]{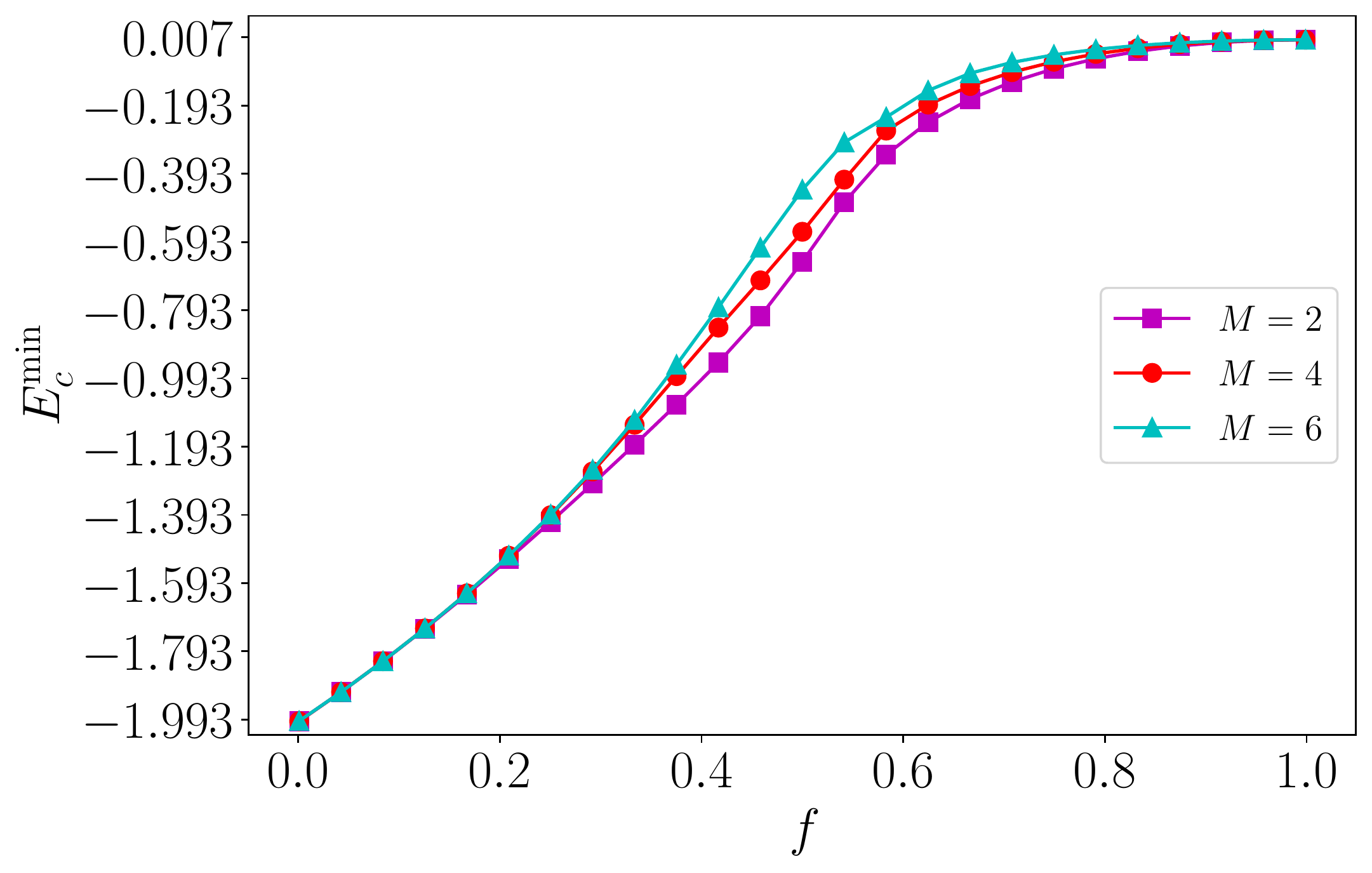}
\includegraphics[width=6cm]{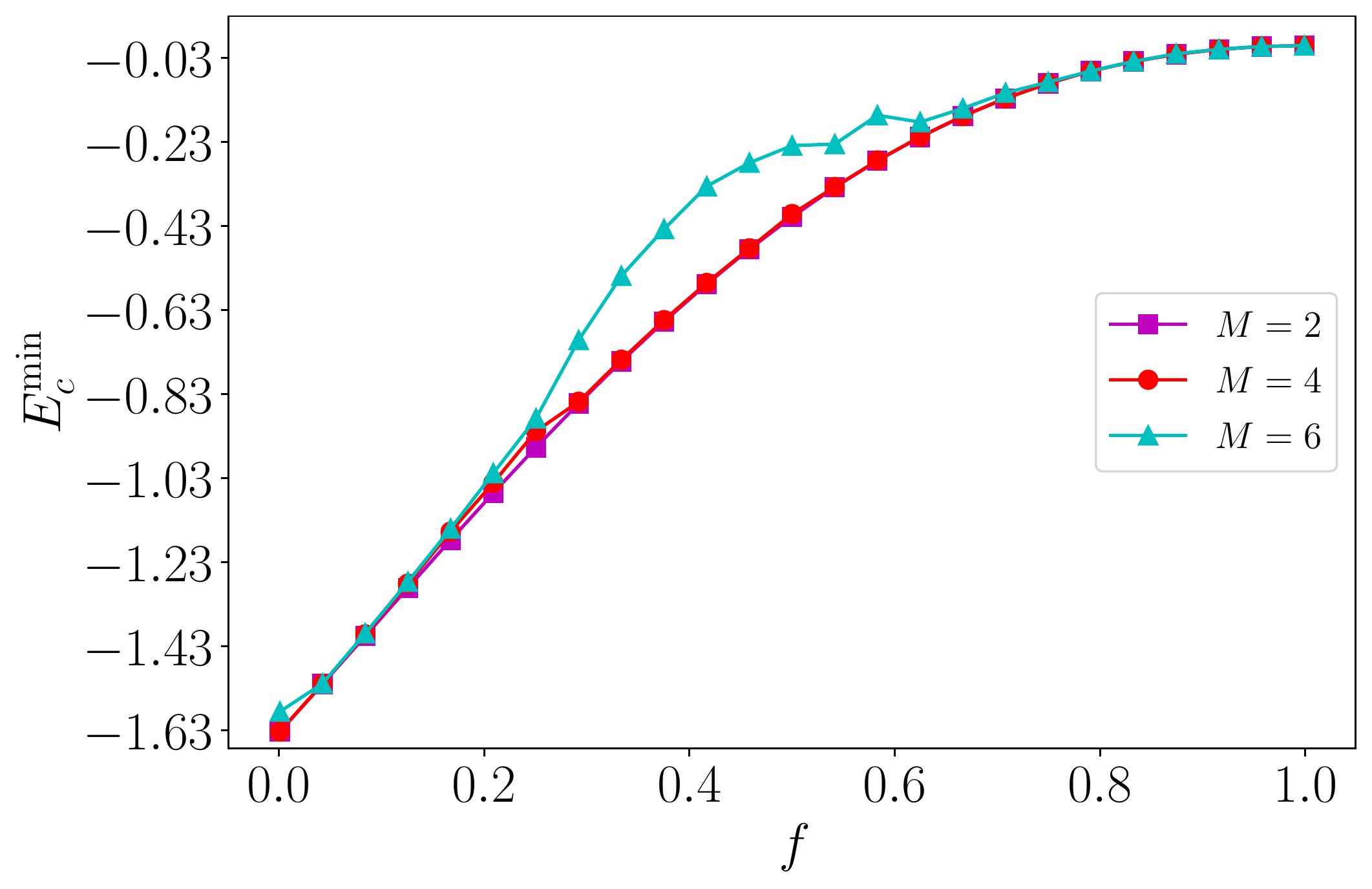}
\includegraphics[width=6cm]{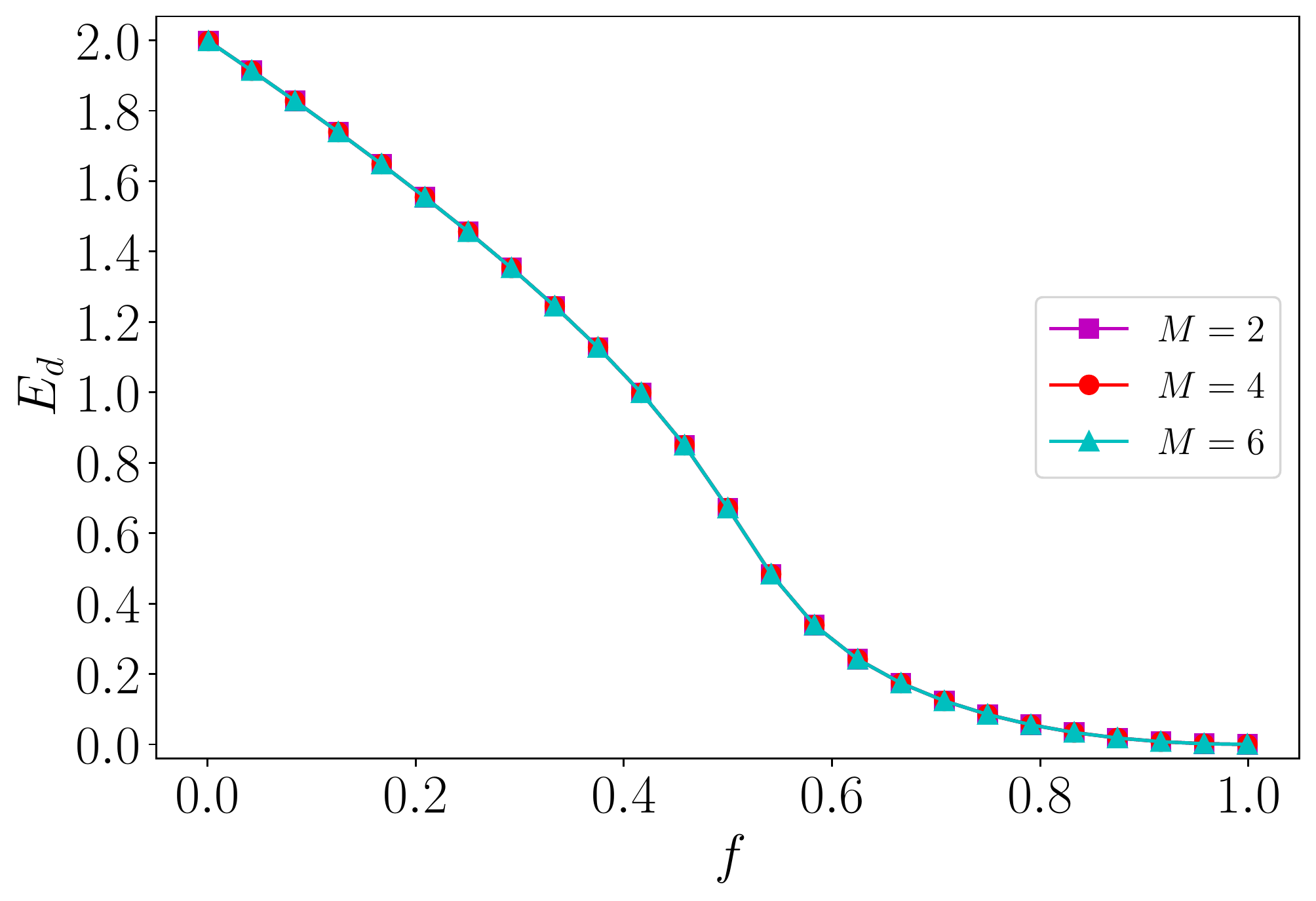}
\includegraphics[width=6cm]{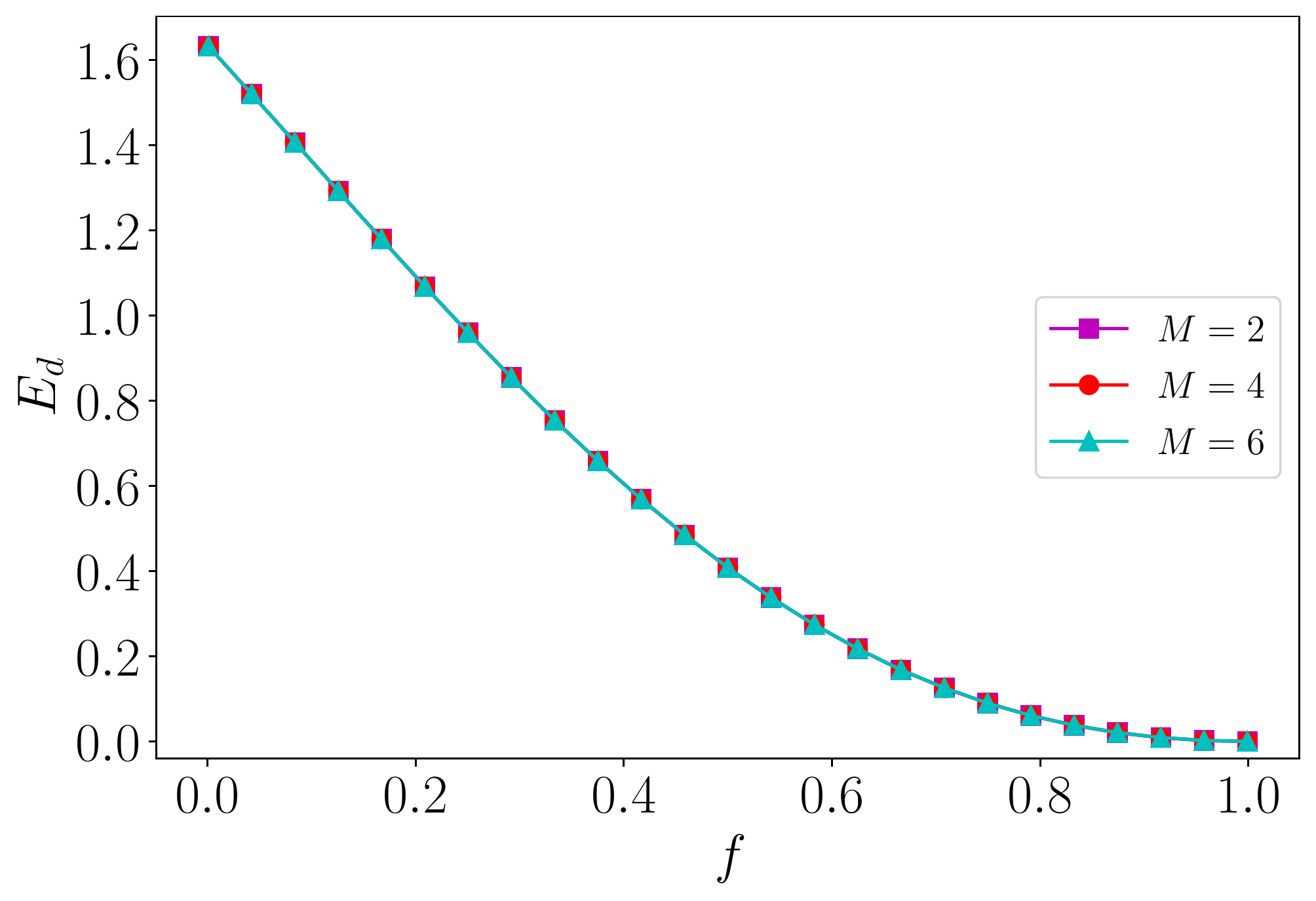}
\includegraphics[width=6cm]{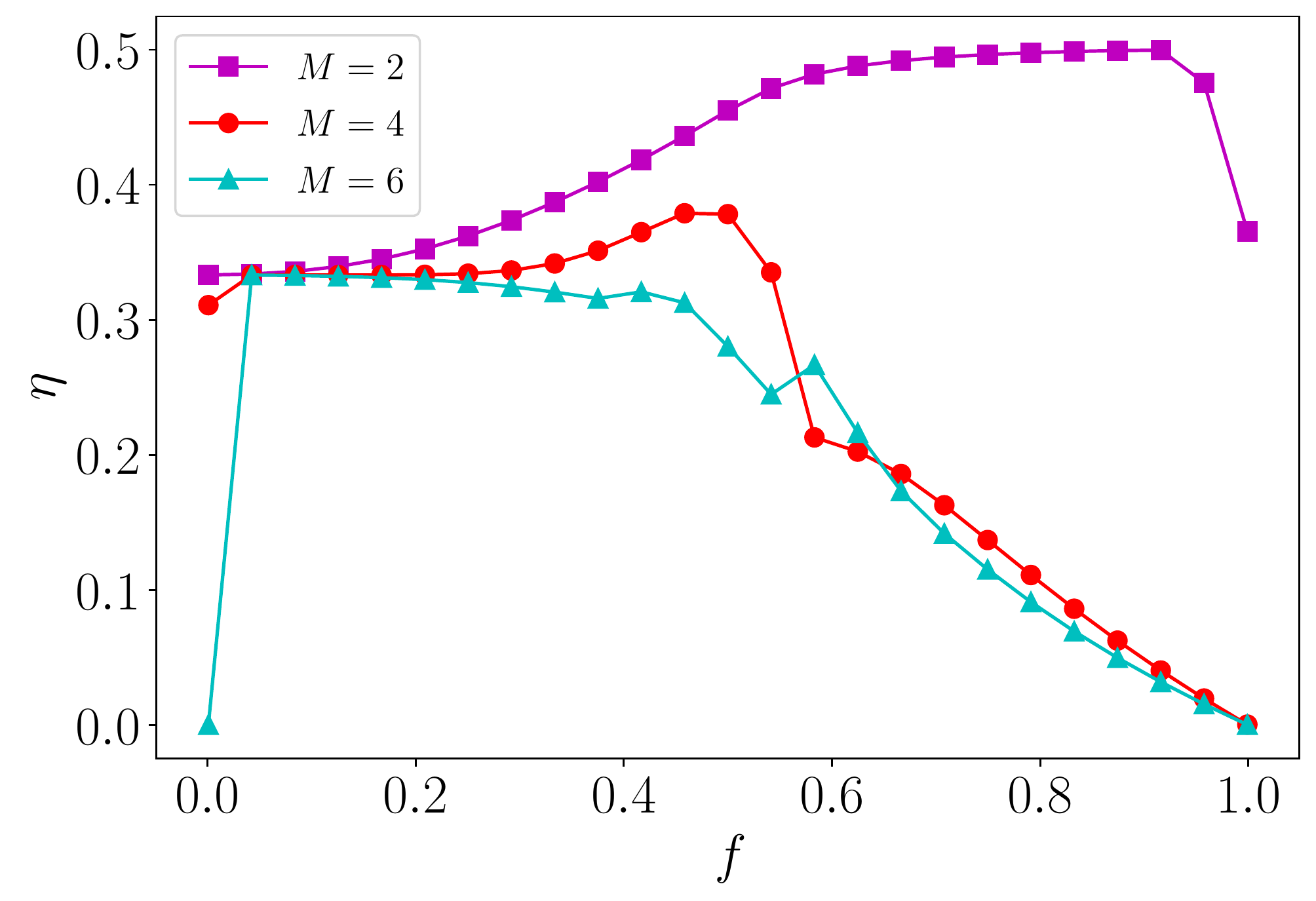}
\includegraphics[width=6cm]{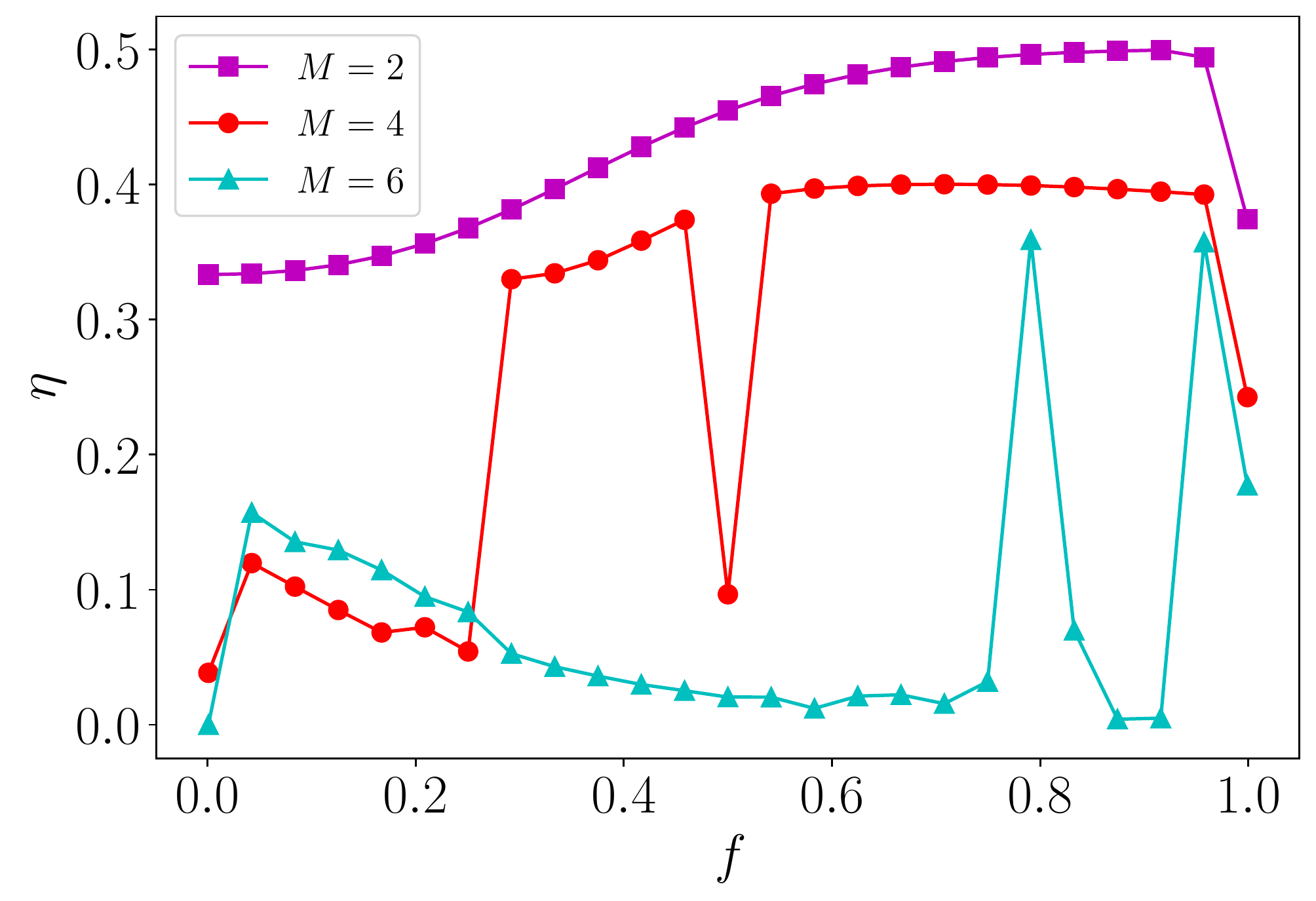}

\caption{Figures of merit of all possible battery--charger configurations of an $8$-node quantum Ising chain with periodic boundary condition. The first column is for $T=0.1$ and the second is for $T=1$. $M$ is the size of the battery. In the last row the efficiency is calculated as $\eta=\mathcal{E}/(E_d+E_c^{\min})$. The abrupt jumps in the efficiency are likely to be related to the error in the multiparameter optimization that is complicated by the presence of multiple local optima (the target function is a sum of cosines, and the optimization is carried over 16 angles for $M=4$ and 64 angles for $M=6$; see Eq.~\eqref{eq:ecminta}.}
\label{fig:Ec}
\end{figure}

Note that, once $M \geq 2$, the amount of $(\alpha < \gamma)$ pairs is strictly larger than the amount of $\alpha$'s, therefore, it will not generally be possible to choose $\vec{\theta}$ so that $\min \sum_{\alpha < \gamma} A_{\alpha, \gamma} \cos (\theta_\alpha - \theta_\gamma + \phi_{\alpha, \gamma}) = - \sum_{\alpha < \gamma} A_{\alpha, \gamma}$.
The numerical minimization can be  carried out by the ``differential evolution'' method, e.g., in \textsc{Python}. The results for the ergotropy, the disconnecting energy, the minimal connecting energy and the efficiency  for a finite chain with $N=8$ are shown in Fig.~\ref{fig:Ec}.

In the first row of Fig.~\ref{fig:Ec} we can see that the ergotropy is still maximum around the value $f_c=1/2$, but it is non-negative for $f>f_c=1/2$, because the reduced state of the disconnected $M$ spins is ``charged'' (or active). 
The bumps on the curves for $E_c^{\min}$ for $T=1$, $M=4,\, 6$, signal possible numerical errors in the multiparameter optimization ($2^M$ angles $\theta_\alpha$).
Furthermore, as $f$ approaches and exceeds 1/2 the sum $E_c^{\min}+E_d$ becomes quite small. These combined factors result in rather imprecise curves for the efficiency, in particular for $M=4$ and $6$.

At least for $M=2$, the efficiency becomes large in the $f \to 1$ regime. This is the weak coupling limit in our model, and, along with the ergotropy, also $E_d$ and $E_c$ vanish, maintaining a finite ratio in Eq.~\eqref{eta:def}. The maximization of the efficiency in the limit of zero output is a type of power--efficiency tradeoff akin to the similar tradeoff for ordinary heat engines \cite{Sekimoto2000, Allahverdyan2013, Shiraishi2016}. In a sense, this limit corresponds to the regime of reversible, quasi-static operation of the device, which, in analogy with the Carnot cycle, is marked by high efficiency at the expense of vanishing output power.

In contrast, Fig.~\ref{fig:Ec} shows that, when the temperature is low ($T = 0.1$), the efficiencies of $4$- and $6$-spin batteries peak around $f = 1/2$, thereby breaking the tradeoff. Although the evidence for this is only partial---the numerical optimization of $E_c$ for $M = 4, \, 6$ could not be carried out precisely enough---we observe in Figs.~\ref{fig:ergo} and \ref{best:DMRG} a similar behavior (efficiency and ergotropy both peaking around $f = 1/2$) for a single-spin battery when the chain is in the ground state $\ket{0^+}$. In both cases, the violation of the power--efficiency tradeoff is due to the proximity of the system to criticality. In this context, it is worth noting that critical systems are responsible for breaking the power--efficiency tradeoff also in ordinary heat engines \cite{Campisi2016, Abiuso2020}, albeit by a completely different mechanism and in a completely different setting.

\section{Conclusions}
\label{sec:conclu}

In this paper, we have studied a four-stroke thermodynamic cycle representing the operation of a quantum battery and its charger. The total system consisting of the battery and the charger is initially either in a ground state or in a thermal equilibrium state, thus protecting the battery's charged state, which is only accessible when disconnected from the charger. 

Here we have expanded our previous work \cite{Hovhannisyan2020} by considering a fully coherent manipulation in the first three strokes of the cycle. Moreover, we consider a battery--charger system that exhibits a quantum phase transition. We have shown that the phases of the eigenstates of the battery's reduced state can be manipulated by the energy extracting protocol to increase the efficiency of the process without compromising the extracted energy. This aspect highlights an important general point that, when manipulating a subsystem of a strongly interacting system, locally irrelevant phases may have a nontrivial effect on the global energetics of the system. This is a purely quantum effect brought about by correlations and noncommutativity.

By operating the working fluid at the verge of the quantum phase transition, all the figures of merits of the device can be further increased. In particular, we found that if the charger-battery is in the ground state, the single spin device only works in the ordered phase, and the critical exponents characterizing the phase transition manifest in the properties of the device close to criticality. Moreover, the arbitrary phase $\theta$ of the unitary that extract the ergotropy can change the scaling exponent of the efficiency from $2\beta$ to a value close to $\beta/2$.

For a battery with $M = 2$ and initially in the thermal ground state $(\ket{0^+} \bra{0^+} + \ket{0^-} \bra{0^-}) / 2$, the ergotropy does not present a critical behavior akin to Eq.~\eqref{erg:crit}. However, the ergotropy does show a special behavior: $d \mathcal{E} / d f$ diverges at the critical point; see \ref{appendixM2}. When the ground state is pure (e.g., $\ket{0^+}$), the critical behavior of $\av{\sigma_i^x}$ given by Eq.~\eqref{sigmax}, and the expected critical behavior of $\av{\sigma_i^x \sigma_{i+1}^z}$ suggested by Fig.~\ref{fig:sigmaxz}b, imply that any expectation value calculated on the reduced state \eqref{uglystate} will have a discontinuous derivative at $f_c$, also signaling a special behavior at criticality.
 
When the charger--battery system is in a thermal state, a minimum of two spins are needed for the battery to deliver energy. As this size increases, the optimization of the relative phases to increase the efficiency becomes nontrivial.

Overall, our results highlight that collective phenomena can be fruitfully exploited in order to enhance the thermodynamic performance of quantum many-body devices.

\section*{Acknowledgements}

F. B. thanks Fondecyt project 1191441 and the Millennium Nucleus ``Physics of active matter'' of ANID (Chile). K. V. H. acknowledges support by the University of Potsdam startup funds.

\appendix

\section{Magnetization and nearest neighbors correlation functions at finite $T$}
\label{App:ChainfiniteT}

At finite temperature, the magnetization and near neighbors correlation functions
needed for the analytical computations of the single and double spin batteries are \cite{Osborne2002,Barouch1970}
\bea
\av{\sigma_i^x}&=&0,\\ 
\av{\sigma_i^y}&=&0,\\
\av{\sigma_i^z}&=&
\frac{1}{\pi}\int_0^\pi d\phi\frac{1+\lambda \cos\phi}{\omega_\phi}\tanh\left(\frac{\omega_\phi}{2k_BT}\right),
\eea 
and
\bea
\av{\sigma_i^x\sigma_{i+1}^y}&=& 0\\
\av{\sigma_i^y\sigma_{i+1}^z}&=& 0\\
\av{\sigma_i^x\sigma_{i+1}^z}&=& 0\\
\av{\sigma_i^x\sigma_{i+1}^x}&=&\frac{1}{\pi}\int_0^\pi d\phi\frac{\cos\phi+\lambda}{\omega_\phi}\tanh\left(\frac{\omega_\phi}{2k_BT}\right),\\
\av{\sigma_i^y\sigma_{i+1}^y}&=& \frac{1}{\pi}\int_0^\pi d\phi\frac{\cos\phi+\lambda\cos 2\phi}{\omega_\phi}\tanh\left(\frac{\omega_\phi}{2k_BT}\right),\\
\av{\sigma_i^z\sigma_{i+1}^z}&=&\av{\sigma_i^z}^2-\av{\sigma_i^x\sigma_{i+1}^x}-\av{\sigma_i^y\sigma_{i+1}^y}
\eea
where we have denoted $\omega_\phi=\sqrt{1+\lambda^2+2\lambda\cos\phi}$ and 
$\av{\cdot}=\Tr[\cdot \frac{e^{H/k_BT}}{Z}]$

\section{The unitary operator $U_{\erg}(\theta)$ for the $M=1$ case}
\label{appendixUM1}

Starting from Eq.~\eqref{ugeneral} for $M=1$, i.e., 
\bea
U_{\erg}=e^{i\theta}\ket{e_+}\bra{r_+}+e^{-i\theta}\ket{e_-}\bra{r_-},
\label{Using}
\eea
 where the eigenkets $\ket{e_{\pm}}=\ket{\pm}$ of $\sigma^z$, are also eigenvectors of the single spin Hamiltonian $H_s=-f \sigma_0^z$ with eigenvales $e_+<e_-$, and where $\ket{r_\pm}$ are the eigenvectors of the reduced state matrix $\rho_{\rm II}$ given in Eq.~(\ref{rho0:eq}), with eigenvalues $r_+>r_-$, and where we neglected an irrelevant global phase.
The eigenvalues of the single spin exhausted state $\rho_{\rm III}$ given in Eq.~(\ref{rho01}) are thus  $r_1=r_+$ and $r_2=r_-$ as given by 
\bea
r_\pm=\frac{1}{2}(1\pm \bsz)
\eea
and the eigenvectors read :
\bea
\ket{r_+}=\cos\alpha\ket{+}+\sin\alpha\ket{-}\\
\ket{r_-}=-\sin\alpha\ket{+}+\cos\alpha\ket{-}
\eea
where  the angle $\alpha$ is implicitly defined through 
\bea
\label{Eq19b}
\cos\alpha=\frac{\av{\sigma^x}}{\sqrt{2\bsz(\bsz-\av{\sigma_0^z})}}\\
\sin\alpha=\frac{\bsz-\av{\sigma_0^z}}{\sqrt{2\bsz(\bsz-\av{\sigma_0^z})}}
\label{Eq20b}
\eea
which are equivalent to 
Eqs.\eqref{Eq19}--\eqref{Eq20}.
Thus from Eq.~(\ref{Using}) one finally finds
\begin{eqnarray}
U_{\erg}(\theta)&=&e^{i\theta}\cos\alpha\ket{+}\bra{+}+e^{i\theta}\sin\alpha\ket{+}\bra{-}-e^{-i\theta}\sin\alpha\ket{-}\bra{+}+e^{-i\theta}\cos\alpha\ket{-}\bra{-}\nonumber \\
&=& e^{i\theta \sigma_0^z}e^{i\alpha \sigma_0^y}.
\end{eqnarray}

\section{Additional information on the numerical results}
\label{num:app}
\begin{figure}[h]
\center
\psfrag{ }[ct][ct][1.]{ }
\includegraphics[width=7.5cm]{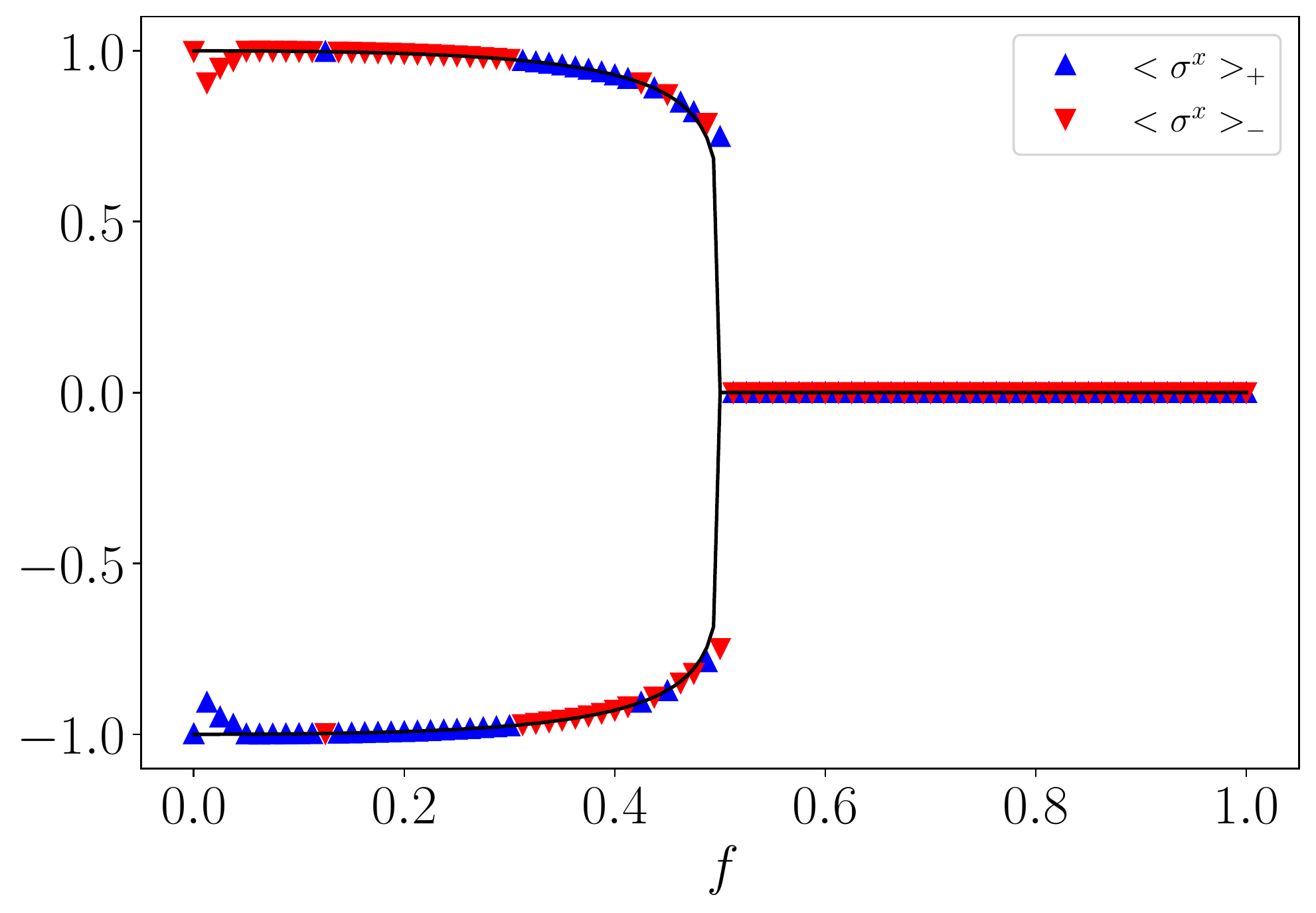}
\includegraphics[width=7.5cm]{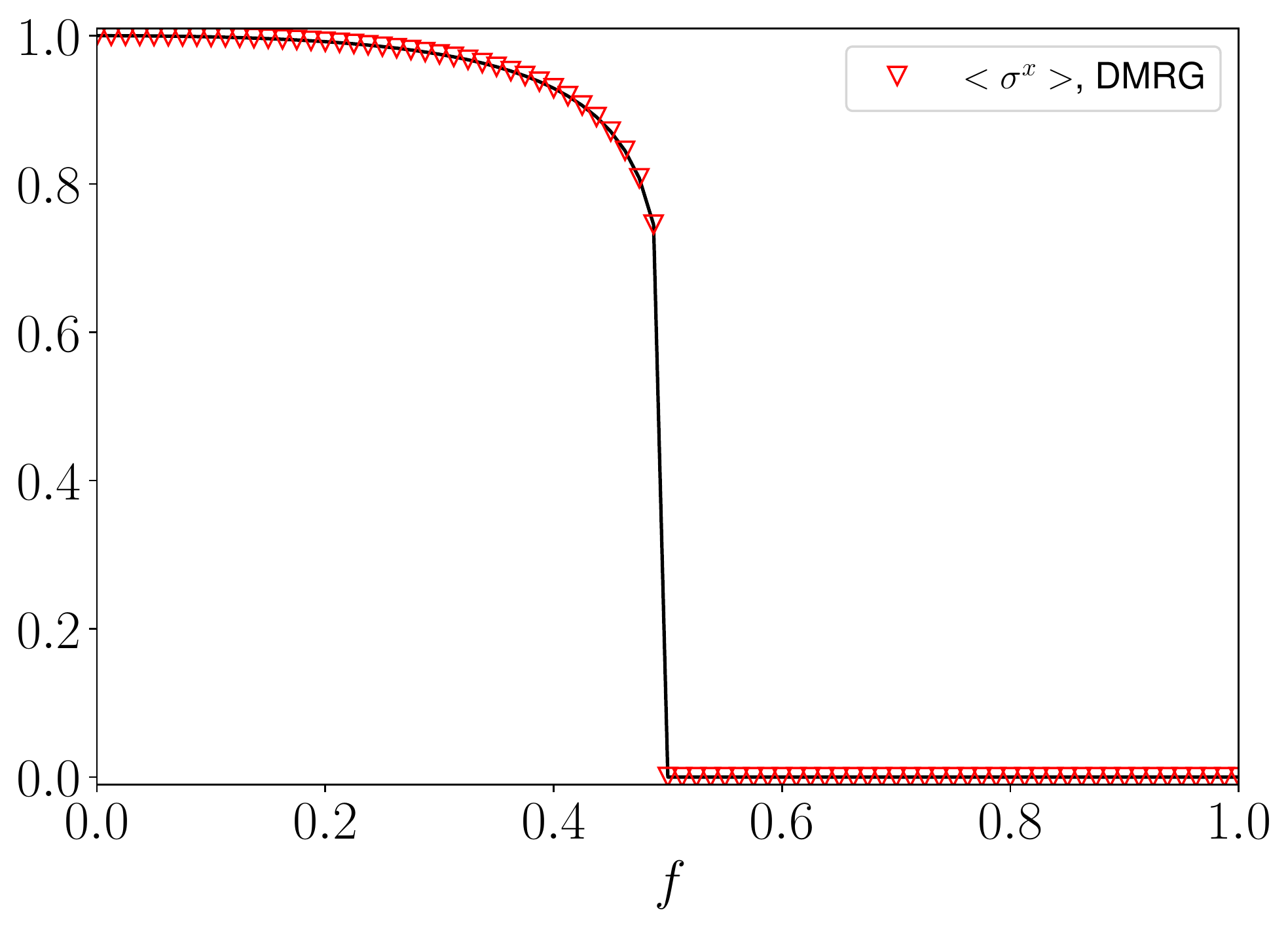}
\caption{Left: magnetization $\av{\sigma^x_i}$ in the ground state of (\ref{HN:def}) as a function of $f$. Full line: exact result  eq.~(\ref{sigmax}). Symbols:  expectation values in the two approximated ground states $\ket{\psi_\pm}$, eqs.~(\ref{psip})--(\ref{psim}). We see that the numerical results ``jump'' between the two brances of the exact solution, corresponding to positive and negative magnetization. This highlights the fact that the two states eqs.~(\ref{psip})--(\ref{psim}) represent a good approximation of the ground state, up to a phase $\pm 1$. Right: longitudinal magnetization $\av{\sigma^x_i}$ in the ground state of (\ref{HN:def}) as a function of $f$. Full line: exact result  eq.~(\ref{sigmax}). Symbols:  expectation value as obtained with the DMRG algorithm.}
\label{fig:sigma}
\end{figure}

In order to obtain the numerical results shown in section \ref{sec:M1}, we use two different approaches. In the first approach we diagonalise the Hamiltonian \eqref{HN:def} for a finite value of $N$ spins, so as to find its ground state.
Technically speaking, the ground state becomes doubly degenerate below $f_c$ only in the thermodynamic limit $N\to \infty$. Thus for any finite $N$ the ground state exhibits vanishing magnetization $ \av{\sigma_i^x}$ even for $f<f_c$, with the energy gap between the ground state and the first excited eigenstate closing in that limit. We resort to the following approximation to evaluate the  two degenerate ground states: let $\ket 0_{N}$ and $\ket 1_{N}$ be the ground state and the first excited state obtained numerically for finite $N$, respectively.
We introduce  
\begin{eqnarray}
\ket{\psi_+}&=&(\ket 0_{N}+\Theta(fc-f)\cdot \ket 1_{N})/\sqrt{2}\label{psip}\\
\ket{\psi_-}&=&(\ket 0_{N}-\Theta(fc-f)\cdot   \ket 1_{N})/\sqrt{2}\label{psim},\\
\Theta(x)&=& 1\quad  \mathrm{iff}\,  x>0; \qquad 0 \quad \mathrm{otherwise},
\end{eqnarray} 
where the kets $\ket{\psi_\pm}$ are our approximation for the two degenerate ground states.
We then compare the exact result for the magnetization $\av{\sigma_i^x}$, eq.~(\ref{sigmax}), with the corresponding expectation values in the two states $\ket{\psi_\pm}$: the results are shown in the left panel of Fig.~\ref{fig:sigma}. We find an acceptable agreement that worsens as $f\to f_c^-$. In the right panel of the same figure we show the results for the magnetization obtained with the other numerical techinque employed in the present paper: namely the DMRG method \cite{Vidal2007, Schollwock2011}.
The DMRG method is a very efficient variational technique  for the simulation of the static and dynamic properties of one-dimensional  quantum lattice systems of the type \eqref{HN:def}.  In particular it is widely used to study the ground state of  such systems.

In this paper we have used the open-source code made available by the OpenMPS project \cite{Jaschke2018, Wall2012} to study the ground state of  \eqref{HN:def}.
 In our DMRG simulations we have taken $N=480$ spins with bond dimension 20.
Inspection of the right panel in Fig.~\ref{fig:sigma} indicates that for such a choice of simulation parameters the agreement with the expected magnetization is excellent, even in proximity of  the critical region. In the  DMRG simulations we have added a term $-h\sum_{i=0}^N \sigma_i^x$ to the system Hamiltonian, with $h=+10^{-9}$ so as to break the symmetry in the ground state, and select the $\ket{0^+}$ state. 

\begin{figure}[h]
\center
\psfrag{ }[ct][ct][1.]{ }
\includegraphics[width=7.5cm]{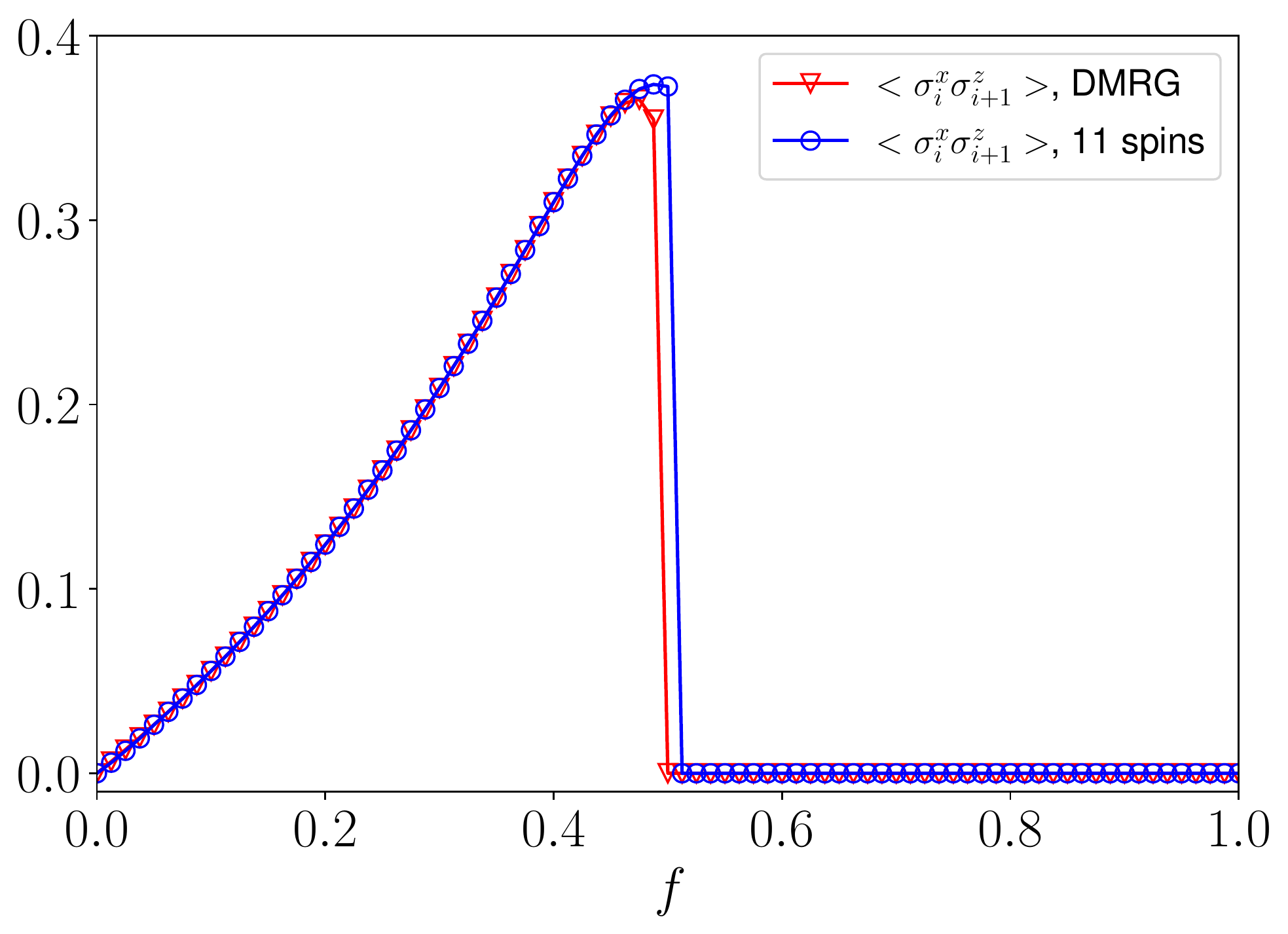}
\includegraphics[width=7.5cm]{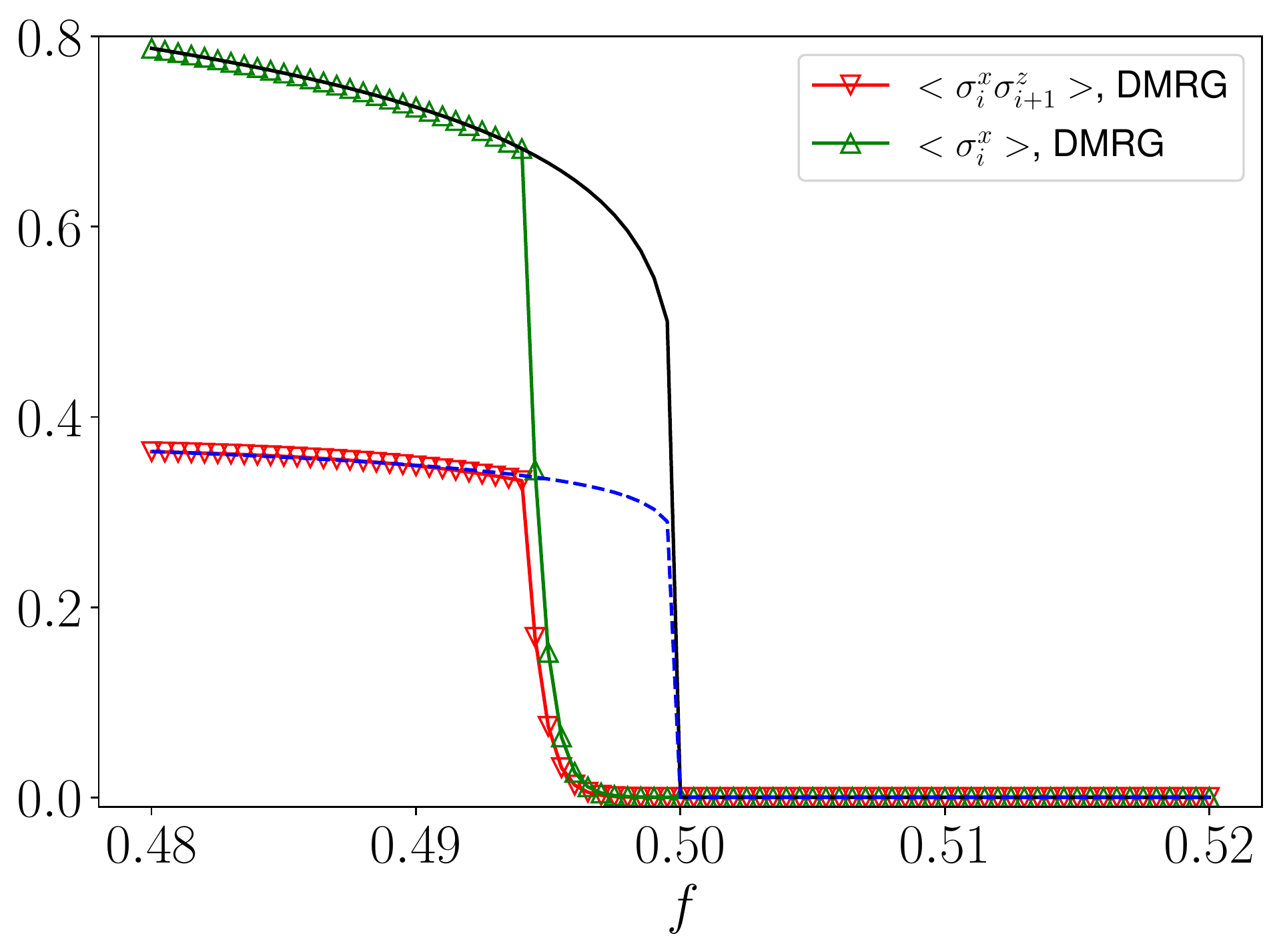}
\caption{Left: correlation  $\av{\sigma^x_i\sigma^z_{i+1}}$ in the ground state of (\ref{HN:def}) as a function of $f$. Right: zoom of the plot  in the critical region. The magnetization  $\av{\sigma^x_i}$ is also plotted for comparison. The full line corresponds to the analytical expression (\ref{sigmax}) for the magnetization. The dashed line is a fit to the data with a function $\sim (1-\lambda^{-2})^{\delta}$ with $\delta=\beta/2$.}
\label{fig:sigmaxz}
\end{figure}

As discussed in the main text the reason for using numerical techniques to find the ground state of the Hamiltonian (\ref{HN:def}) lies in the fact that the correlation $\av{\sigma^x_i\sigma^z_{i+1}}$ has no explicit expression in terms of the system parameters. Such a correlation is required in order to calculate the connecting energy, Eq.~\eqref{Wc:defp}. In Fig.~\ref{fig:sigmaxz} we plot the correlation $\av{\sigma^x_i\sigma^z_{i+1}}$ as a function of $f$, as obtained with the two numerical approaches discussed above. By zooming in on the critical region, we see that the DMRG method gives a vanishing longitudinal magnetization and correlation even for $f$ smaller than $f_c$, while one would expect  $\av{\sigma^x_i}\neq 0$ below $f_c$. This is  clearly due to the failure of the algorithm: at  the critical point the system becomes infinitely correlated, so any numerical method based on finite size approximation will fail as the critical point is approached. In the right panel of Fig.~\ref{fig:sigmaxz} we also propose a  power law for the correlation based on a fit to the numerical data.

\section{Ergotropy of two sites}
\label{appendixM2}

We are interested in evaluating the ergotropy of two spins (sites $i=0$ and $1$ in the following), disconnected form the chain ($M=2$ in Fig.~\ref{fig:cycle}). In this case the Hamiltonian of the system is
\[
H'_S=H_S/f=-\lambda \sigma^x_0\sigma^x_1-\sigma^z_0-\sigma^z_1
=\left(\begin{array}{cccc}
-2&0&0&-\lambda\\
0&0&-\lambda&0\\
0&-\lambda&0&0\\
-\lambda&0&0&2
\end{array}\right),
\]
with the $4\times4$ matrix being its representation in the canonical basis ($++,+-,-+,--$).
 The eigenvalues and eigenvectors are
\bea
\epsilon_1^\uparrow&=&-\sqrt{4+\lambda^2},\quad \ket{\epsilon_1^\uparrow}=(\sin\mu_1,0,0,\cos\mu_1)\\
\epsilon_2^\uparrow&=&-\lambda,\quad \ket{\epsilon_2^\uparrow}=\frac{1}{\sqrt{2}}(0,1,1,0)\\
\epsilon_3^\uparrow&=&\lambda,\quad \ket{\epsilon_3^\uparrow}=\frac{1}{\sqrt{2}}(0,-1,1,0)\\
\epsilon_4^\uparrow&=&\sqrt{4+\lambda^2},\quad \ket{\epsilon_4^\uparrow}=(\sin\mu_4,0,0,\cos\mu_4)=(\cos\mu_1,0,0,-\sin\mu_1)
\eea

\[
\tan\mu_i=\frac{2-\epsilon_i^\uparrow}{\lambda}\Rightarrow \mu_4=\mu_1+\frac{\pi}{2}
\]

If the state $\varrho_{\rm I}$ of the spin chain is thermal, $\av{\sigma_0^x}=\av{\sigma_0^y}=0$ and $\av{\sigma_0^\zeta\sigma_1^\gamma}=0$ if $\zeta\neq \gamma$. The reduced state of the $M=2$ spins is 
\begin{eqnarray}
\rho_2&=&\frac{1}{4}\left(I_4+\av{\sigma_0^z}(\sigma^z_0+\sigma^z_1)+\sum_{\gamma=x,y,z}\av{\sigma_0^\gamma\sigma_1^\gamma}\sigma_0^\gamma\sigma_1^\gamma\right)\nonumber \\
&=&\frac{1}{4}\left(
\begin{array}{cccc}
\Omega_+&0&0&\Delta\\
0&1-\av{\sigma^z\sigma^z}&\av{\sigma^x\sigma^x}+\av{\sigma^y\sigma^y}&0\\
0&\av{\sigma^x\sigma^x}+\av{\sigma^y\sigma^y}&1-\av{\sigma^z\sigma^z}&0\\
\Delta &0&0&\Omega_-
\end{array}\right),
\label{eq41}
\end{eqnarray} 
with
\bea
\Omega_+=1+\av{\sigma_0^z\sigma_1^z}+2\av{\sigma_0^z},\\
\Omega_-=1+\av{\sigma_0^z\sigma_1^z}-2\av{\sigma_0^z},\\
\Delta=\av{\sigma_0^x\sigma_1^x}-\av{\sigma_0^y\sigma_1^y}.
\eea

The magnetization and correlations were computed in \cite{Barouch1970}. 
The eigenvalues and eigenvectors of $\rho_2$ are
\bea
r_1^\downarrow&=&\frac{1}{4}(1+\sqrt{\Delta^2+4\av{\sigma_0^z}^2}+\av{\sigma_0^z\sigma_1^z}),\quad \ket{r_1^\downarrow}=(\sin\nu_1,0,0,\cos\nu_1),\\
r_2^\downarrow&=&\frac{1}{4}(1+|\av{\sigma_0^x\sigma_1^x}+\av{\sigma_0^y\sigma_1^y}|-\av{\sigma_0^z\sigma_1^z}),\quad \ket{r_2^\downarrow}=\frac{1}{\sqrt{2}}(0,1,1,0),\\
r_3^\downarrow&=&\frac{1}{4}(1-|\av{\sigma_0^x\sigma_1^x}+\av{\sigma_0^y\sigma_1^y}|-\av{\sigma_0^z\sigma_1^z}),\quad \ket{r_3^\downarrow}=\frac{1}{\sqrt{2}}(0,-1,1,0),\\
r_4^\downarrow&=&\frac{1}{4}(1-\sqrt{\Delta^2+4\av{\sigma_0^z}^2}+\av{\sigma_0^z\sigma_1^z}),\nonumber \\
\ket{r_4^\downarrow}&=&(\sin\nu_4,0,0,\cos\nu_4)=(\cos\nu_1,0,0,-\sin\nu_1),
\eea
with 
\bea
\tan\nu_1=
\frac{2\av{\sigma_0^z}+\sqrt{\Delta^2+4\av{\sigma_0^z}^2}} {\Delta},\\
\tan\nu_4=
\frac{2\av{\sigma_0^z}-\sqrt{\Delta^2+4\av{\sigma_0^z}^2}}{\Delta},\\
\nu_4=\nu_1+\frac{\pi}{2}.
\eea
The explicit expression of the ergotropy extraction unitary $U_{\erg}(\{\theta\})=\sum_{j=1}^4 e^{i\theta_j}\ket{\epsilon_j^\uparrow}\bra{r_j^\downarrow}$ is quite involved but it has two $2\times2$ independent blocks, one in the subspace associated to $+-,-+$ and the other to $++,--$. The former is
 a trivial block because $\ket{\epsilon_{2,3}^\uparrow}=\ket{r_{2,3}^\downarrow}$ are independent of $\lambda$. 
 There are three independent phases in $U_{\erg}(\{\theta\}).$
In fact, factorizing a global phase $e^{\frac{i}{2}(\theta_2+\theta_3)}$ and defining 
$\vartheta_{1,4}=\theta_{1,4}-\frac{\theta_2+\theta_3}{2},\vartheta\equiv\frac{\theta_2-\theta_3}{2}$ we have
\bea
U_{\erg}(\{\theta\})=
\left(
\begin{array}{cccc}
a_{11}e^{i\vartheta_1}+a_{44}e^{i\vartheta_4}&0&0&a_{14}e^{i\vartheta_1}-a_{41}e^{i\vartheta_4}\\
0&\cos\vartheta&i\sin\vartheta&0\\
0&i\sin\vartheta&\cos\vartheta&0\\
a_{41}e^{i\vartheta_1}-a_{14}e^{i\vartheta_4}&0&0&a_{44}e^{i\vartheta_1}+a_{11}e^{i\vartheta_4}
\end{array}
\right)
\label{eq48p}
\eea
with
\bea
a_{11}=\sin\mu_1\sin\nu_1, \quad
a_{14}=\sin\mu_1\cos\nu_1,\quad
a_{41}=\cos\mu_1\sin\nu_1,\quad
a_{44}=\cos\mu_1\cos\nu_1.
\eea

As discussed in the main text, to compute the ergotropy, we can consider that all the phases vanishes.
Because $H'_S$, $\rho_2$ and $U\equiv U_{\erg}(\{0\})$ have all the same block structure, and because the action of $U$ in the $+-,-+$ space is trivial (the difference $\rho_2-U\rho_2 U^\dag$ in the block $+-,-+$ vanishes), the ergotropy is only associated to the $++,--$ block, i.e., 
\[
\erg/f=\Tr[h_S(r-uru^\dag)]
\]
with
\bea
h_S=\left(\begin{array}{cc}
-2&-\lambda\\
-\lambda&2
\end{array}\right),\quad
r=\frac{1}{4}\left(
\begin{array}{cc}
\Omega_+&\Delta\\
\Delta &\Omega_-
\end{array}\right),\quad
u=\left(
\begin{array}{cc}
\cos(\delta)&\sin(\delta)\\
-\sin(\delta)&\cos(\delta)
\end{array}
\right)
\nonumber
\eea
with  $\delta=\mu_1-\nu_1$, so we obtain
\bea
\erg/f=( \Delta- \lambda  \av{\sigma_0^z})\sin2\delta-
(\Delta  \lambda +4\av{\sigma_0^z}) \sin^2\delta.
\label{ergMigual2}
\eea
\begin{figure}[h]
\center
\includegraphics[width=8cm]{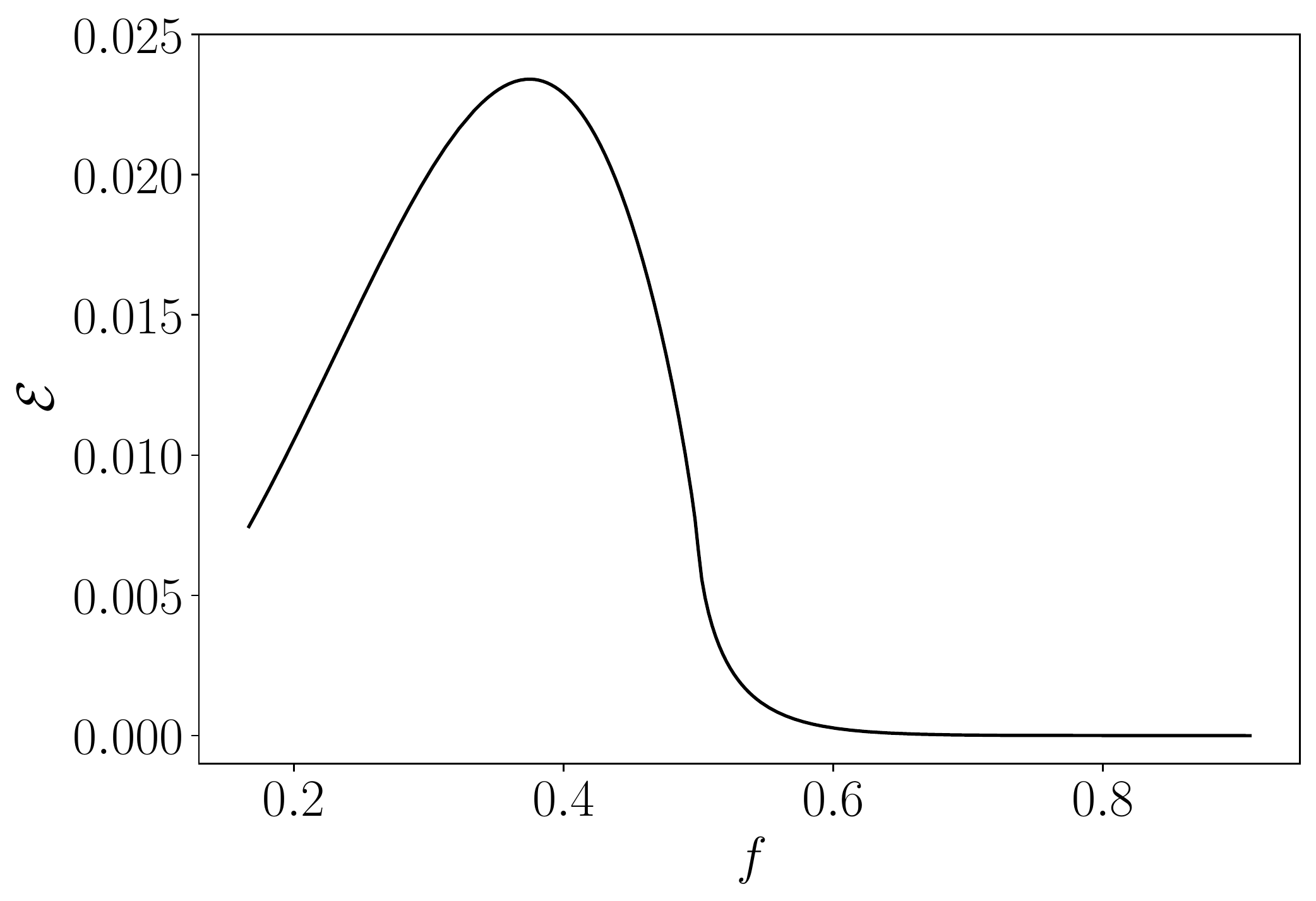}
\caption{
Ergotropy (\ref{ergMigual2}) of the two spin battery ($M=2$) as a function of $f=1/(\lambda+1)$, for the infinite chain at $T=0$.}
\end{figure}

The ergotropy exists even in the thermal ground state at $T=0$, i.e.,  $(\ket{0^+}\bra{0^+}+\ket{0^-}\bra{0^-})/2.$ Note that $\Delta$ is not differentiable at $f_c$ ~\cite{Pfeuty1970}, in fact the derivative diverges logarithmically ($d \Delta / d f \propto \ln |f - f_c|$), thus although different to what we expect for the ergotropy in the state $\ket{0^+}\bra{0^+}$, the thermal ground state does display critical behavior.

In the ``symmetry broken'' ground state $\ket{0^+}$ the subsystem with $M=2$ has for reduced state
\begin{equation}
\rho_2=\frac{1}{4}\left(I_4+\sum_{\mu=0}^1(\av{\sigma_0^x}\sigma_\mu^x+\av{\sigma_0^z}\sigma^z_\mu)+\sum_{\gamma,\zeta=x,y,z}\av{\sigma^\gamma_0\sigma^\zeta_1}\sigma^\gamma_0\sigma^\zeta_1\right),
\label{uglystate}
\end{equation} 
which have additional non-vanishing terms $\av{\sigma_0^x}\neq 0$ and $\av{\sigma_0^x\sigma_1^z}\neq 0$ for $\lambda>1$ destroying the X shape 
structure of the reduced thermal state [cf. eq~(\ref{eq41})] that simplified the previous calculation.

\section*{References}

\bibliographystyle{iopart-num}

\bibliography{bibliography}

\end{document}